\documentclass[12pt]{iopart}

\pdfoutput=1

\usepackage{tgtermes}

\usepackage{float}
\usepackage{graphicx}  
\usepackage{xcolor}
\usepackage{bm}        
\usepackage{braket}
\usepackage{amssymb}   
\usepackage{epstopdf}
\usepackage{xfrac}
\usepackage{subfigure}
\usepackage{anyfontsize}

\usepackage{pifont} 

\usepackage{amsmath}
\usepackage{amsfonts}
\usepackage{bbm}
\definecolor{darkblue}{rgb}{0.1,0.2,0.6}
\definecolor{darkred}{rgb}{0.8,0.1,0.2}
\usepackage[colorlinks,citecolor=darkblue,linkcolor=darkred,urlcolor=darkblue]{hyperref} 
\usepackage[margin=30pt, bf, font=footnotesize, center, justification=justified]{caption}[2004/07/16]
\usepackage{tikz}
\usepackage{enumerate}
\usepackage{setspace}
\usepackage{url}  
\usepackage{mathrsfs}

\usepackage{etoolbox}

\usepackage{lipsum}

\makeatletter
\def\@mkboth#1#2{}
\newlength\appendixwidth
\preto\appendix{\addtocontents{toc}{\protect\patchl@section}}
\newcommand{\patchl@section}{%
  \settowidth{\appendixwidth}{\textbf{Appendix }}%
  \addtolength{\appendixwidth}{1.5em}%
  \patchcmd{\l@section}{1.5em}{\appendixwidth}{}{\ddt}%
}
\makeatother

\newcommand{\sT}{\sf T}

\newcommand{\I}{\mathbb{I}}

\newcommand{\sgn}{\text{sign}}

\newcommand{\xv}{\textbf{x}}
\newcommand{\ev}{\textbf{e}}

\catcode`@=11 
\renewcommand\tableofcontents{%
  \section*{\contentsname}%
  \@starttoc{toc}%
}
\catcode`@=12

\usepackage[normalem]{ulem}

\begin{document}
\title{Finite-temperature entanglement negativity of free fermions}
\author{Hassan Shapourian, Shinsei Ryu}
\address{James Franck Institute and Kadanoff Center for Theoretical Physics, University of Chicago, Illinois 60637, USA.}
\date{\today}

\begin{abstract}
The entanglement entropy of free fermions with a Fermi surface is known to obey a logarithmic scaling and violate the area law  in all dimensions.
Here, we would like to see how temperature affects the logarithmic scaling behavior.
To this end, we compute the entanglement negativity of free fermions using the fermionic partial transpose developed in our earlier paper [\href{https://journals.aps.org/prb/abstract/10.1103/PhysRevB.95.165101}{Phys. Rev. B {\bf 95}, 165101 (2017)}]. In one dimension, we analytically derive the leading order term in the finite-temperature entanglement negativity and show how the entanglement negativity indicates a crossover 
from a quantum entangled state to a classical thermal state, where the entanglement is completely lost.
We explain how the one-dimensional result can be generalized to codimension-one Fermi surface of arbitrary shape in higher dimensions. In both one and two dimensions, we check that our analytical results agree with the numerical simulation of free fermions on a  lattice.
\end{abstract}

\maketitle
{\hypersetup{linkcolor=black} \tableofcontents}

\section{Introduction}

An effective way to characterize quantum states of many-body systems is in terms of quantum entanglement shared between two parts of the system. For a given pure state, e.g., the ground state of a Hamiltonian, entanglement entropies (EEs) such as the von Neumann and
the R\'enyi entropies~\cite{Amico_rev2008,Calabrese_intro2009} provide good quantitative measures of the bipartite entanglement. A notable property of EEs is that the leading order term shows certain universal scaling behaviors as a function of system size. In (1+1)d systems, it is well-known that EEs of gapped states assume a boundary law~\cite{Hastings}, i.e., they saturate as the system is made larger, while EEs of critical (bosonic or fermionic) states described by conformal field theories (CFTs) increase logarithmically with system size~\cite{HOLZHEY1994443,Vidal2003,Calabrese2004,Calabrese2009}, i.e., $S\sim \ln L$ where the coefficient is proportional to the central charge. 
In contrast, in higher dimensions, most systems obey a boundary (area) law~\cite{Plenio_rev2010}. However, there are important exceptions to the boundary law in higher dimensions: Namely, free fermions with a Fermi surface~\cite{Wolf2006,Klich2006,Barthel2006,Li2006,Ding2008,Swingle2010,Helling2011,Swingle2012} and Fermi liquids~\cite{Swingle3d_2010,Swingle2010_Renyi}.
In the case of codimension-one Fermi surfaces in
$d$-dimensions, the entanglement entropies to the leading order scale as $S\sim L^{d-1} \ln L$.


Unlike pure states, entanglement entropies are not good measures of the `useful' entanglement in mixed states, e.g. finite temperature states of quantum systems or tripartite entanglement of a pure state.
This should be expected since a generic mixed state contains both quantum and classical correlations.
The former can be used as a resource for quantum computation (i.e.~, useful entanglement), while the latter could be a result of local operations and classical communications (LOCCs) and is not a resource for quantum computation.
In the case of finite temperature density matrices, the classical correlations are due to thermal fluctuations.
At sufficiently high temperatures thermal fluctuations put the system in an equal superposition of all quantum states, which is in essence fully classical.
Given this, it is important to distinguish between classical mixing and quantum entanglement.
 However, it is well-known that the usual bipartite von Neumann or R\'enyi entropies cannot exclusively capture the quantum entanglement. 
 The (logarithmic) entanglement negativity defined in terms of partial transpose of the density matrix is known as a good candidate to detect the quantum entanglement of mixed states~\cite{PlenioEisert1999,Vidal2002,Plenio2005}. From the quantum information perspective, positive partial transpose (PPT) is a necessary condition for  the separability of density matrices~\cite{Peres1996,Horodecki1996,Simon2000,PhysRevLett.86.3658,PhysRevLett.87.167904,Zyczkowski1,Zyczkowski2}.
The effectiveness of entanglement negativity in the study of finite-temperature systems has been already discussed~\cite{Calabrese_Ft2015,Eisler_Neq,Sherman2016}. A remarkable finding is that the logarithmic negativity decays monotonically as the temperature is increased, which implies a cross-over from a quantum entangled state to a classical mixed state.


Furthermore, entanglement negativity has been studied in various systems ranging from
Harmonic oscillator chains~\cite{PhysRevA.66.042327,PhysRevLett.100.080502,PhysRevA.78.012335,Anders2008,PhysRevA.77.062102,PhysRevA.80.012325}, quantum spin chains
 \cite{PhysRevA.80.010304,PhysRevLett.105.187204,PhysRevB.81.064429,PhysRevLett.109.066403,Ruggiero_1,PhysRevA.81.032311,PhysRevA.84.062307,Mbeng} and topologically ordered phases of matter in (2+1)d~\cite{Wen2016_1,Wen2016_2,PhysRevA.88.042319,PhysRevA.88.042318} 
 to holographic models~\cite{Rangamani2014,Chaturvedi_1,Chaturvedi_2,Chaturvedi2018,Malvimat,Jain_1,Jain_2,Jain_3} and other generic states~\cite{Clabrese_network2013,Alba2013,PhysRevB.90.064401,Nobili2015}.
 In the case of (1+1)d CFTs, a systematic approach has been developed in terms of correlators of the twist fields~\cite{Calabrese2012,Calabrese2013,Ruggiero_2,Alba_spectrum}, where a logarithmic dependence on subsystem size was observed similar to the entanglement entropies. This approach was further extended to CFTs at finite temperatures~\cite{Calabrese_Ft2015}.


In this paper, we study the entanglement negativity of free fermions with a Fermi surface. 
In our earlier work~\cite{Shap_pTR}, we have already shown that the zero-temperature entanglement negativity of free fermions in (1+1)d shows a logarithmic violation of the area law. 
Here, we would like to see the fate of this logarithmic term at finite
temperatures and also ask how our result in (1+1)d is generalized to higher
dimensions. In general, we find that there are two temperature regimes:
First, a quantum regime at low temperatures when $L_\beta> L$, where $L_\beta=\hbar v_F/k_B T$ is  the length scale associated with thermal fluctuations (and $v_F$ is the Fermi velocity) and $L$ is the system size,  
and the entanglement negativity remains almost unchanged and close to its value at the zero temperature. Second, a classical regime when $L_\beta \sim L$ and the entanglement negativity decays asymptotically to zero. \textcolor{blue}{The latter regime is characterized by an area-law entanglement of the form $L^{d-1}\ln L_\beta$.}
This behavior is a direct signature of quantum-to-classical crossover at finite temperatures. 
Compared with the general (bosonic) CFT calculations~\cite{Calabrese_Ft2015} in (1+1)d where the negativity at low temperatures $L<L_\beta$ was difficult to be derived, our analytical results cover both regimes of temperatures and faithfully interpolate between them.


Regarding the method used to compute the entanglement negativity of fermions, we should note that the partial transpose of fermionic density matrices (and hence the entanglement negativity) involves some subtleties due to the Fermi statistics which causes fermion operators to anti-commute. Here, we adopt a definition of fermionic partial transpose in terms of partial time-reversal transformation~\cite{Simon2000,Sanpera1997}, which was introduced in our previous works~\cite{Shap_unoriented,Shap_pTR,Shiozaki_antiunitary}. There exists another definition of partial transpose for fermions based on the fermion-boson mapping (Jordan-Wigner transformation)  which was originally proposed in Ref.~\cite{Eisler2015} and was also followed in other studies~\cite{Coser2015_1,Coser2016_1,Coser2016_2,PhysRevB.93.115148,PoYao2016,Herzog2016,Eisert2016}. 
A survey of differences between these two definitions can be found in Refs.~\cite{Shap_pTR} and~\cite{Shap_sep}.

Let us briefly mention a few major differences between our definition and the other definition of partial transpose (which we may refer to as the bosonic partial transpose since it is unitary equivalent to the bosonic partial transpose). 
The bosonic partial transpose~\cite{Eisler2015} does not respect the tensor product structure of fermionic density matrices in fermion-number parity conserving systems and hence the associated entanglement negativity does not satisfy the (sub-)additivity property. Furthermore, there exist fermionic density matrices which are inseparable~\cite{Banuls2007,Banuls2009,Benatti2014} such as in the Kitaev Majorana chain, while the bosonic partial transpose gives a zero negativity.
From practical standpoint, the bosonic partial transpose leads to an unexpected complication that a partially transposed Gaussian state is not necessarily a Gaussian state anymore. This makes the process of finding entanglement negativity very difficult even for simple non-interacting fermionic systems. 
In Refs.~\cite{Shap_pTR} and~\cite{Shap_sep}, we have shown that these issues do not carry over to our definition of partial transpose.






The rest of the paper is organized as follows: In Sec.~\ref{sec:background} we provide some basic definitions and briefly review the fermionic partial transpose. In Sec.~\ref{sec:finT1d}, we use the spacetime picture to derive the entanglement negativity of free fermions in (1+1)d. As a warm-up, we start this section by explaining how a similar calculation is done for more familiar quantities such as the R\'enyi entropies. We also check the agreement between analytical results and numerical simulation of lattice fermions.
We further discuss the generalization of (1+1)d results to higher dimensions in Sec.~\ref{sec:finT2d} and benchmark our formulas in the case of free fermions in (2+1)d on a square lattice.
We give our concluding remarks and future directions in Sec.~\ref{sec:conclusions}. In addition, some details of analytical derivations and numerical calculations are provided in four appendices.





\section{Background and general remarks}
\label{sec:background}

In this section, we briefly review the definition of partial transpose for fermionic density matrices~\cite{Shap_pTR}. Before we explain the fermionic systems, let us begin our discussion by recapitulating the definition of the partial transpose and the corresponding entanglement measure for bosonic systems. Consider two subsystems $A_1$ and $A_2$, where the overall Hilbert space is a tensor product $\mathcal{H}= \mathcal{H}_1\otimes \mathcal{H}_2$.
Let
\begin{align}
\rho = \sum_{ijkl} \braket{e_i^1,e_j^2|\rho |e_k^1,e_l^2} \ket{e_i^1,e_j^2} \bra{e_k^1,e_l^2}, 
\end{align}
be the density matrix expanded in an orthonormal basis $\ket{e_i^1,e_j^2}$ of $\mathcal{H}$,
where $\ket{e_i^{1}}$ and $\ket{e_j^{2}}$ span the $\mathcal{H}_1$ and $\mathcal{H}_2$ Hilbert spaces, respectively. 
The partial transpose with respect to the subsystem $A_1$ is defined by exchanging the matrix elements in that subsystem,
\begin{align}
\label{eq:parttrans_b}
\rho^{\sT_1} := \sum_{ijkl} \braket{e_k^1,e_j^2|\rho |e_i^1,e_l^2} \ket{e_i^1,e_j^2} \bra{e_k^1,e_l^2}.
\end{align}
Equivalently, the partial transpose acting on a basis operator amounts to
\begin{align}
\big( \ket{e_i^1,e_j^2} \bra{e_k^1,e_l^2} \big)^{\sT_1} := \ket{e_k^1,e_j^2} \bra{e_i^1,e_l^2}. 
\label{eq:app_f1}
\end{align}
We should note that the action of the partial transpose in the above definition
follows directly from the usual matrix transposition and this is a mathematically
meaningful operation on bosonic density matrices associated with finite-dimensional Hilbert spaces, because their Hilbert
space admit a tensor product structure where the operator algebra is isomorphic to
finite-dimensional Weyl algebra with positive characteristic (i.e., matrix algebra).
For the time-being, let us recall that this is not the case for fermionic systems, since ordering of the operators (or states) matters due to the anti-commuting property of fermion operators (Clifford algebra). 

The logarithmic negativity is then defined by
\begin{align} \label{eq:neg_def}
{\cal E}:= \ln \Tr |\rho^{\sT_1}|, 
\end{align}
where $\Tr |O|:=\Tr \sqrt{O^\dag O}$ means the sum of the square roots of the eigenvalues of the product operator $O^\dag O$, or in short, the trace norm of the operator $O$. If $O$ is Hermitian, the trace norm will be simplified into the absolute value of the eigenvalues of $O$.  

\begin{figure}
\center
\includegraphics[scale=0.75]{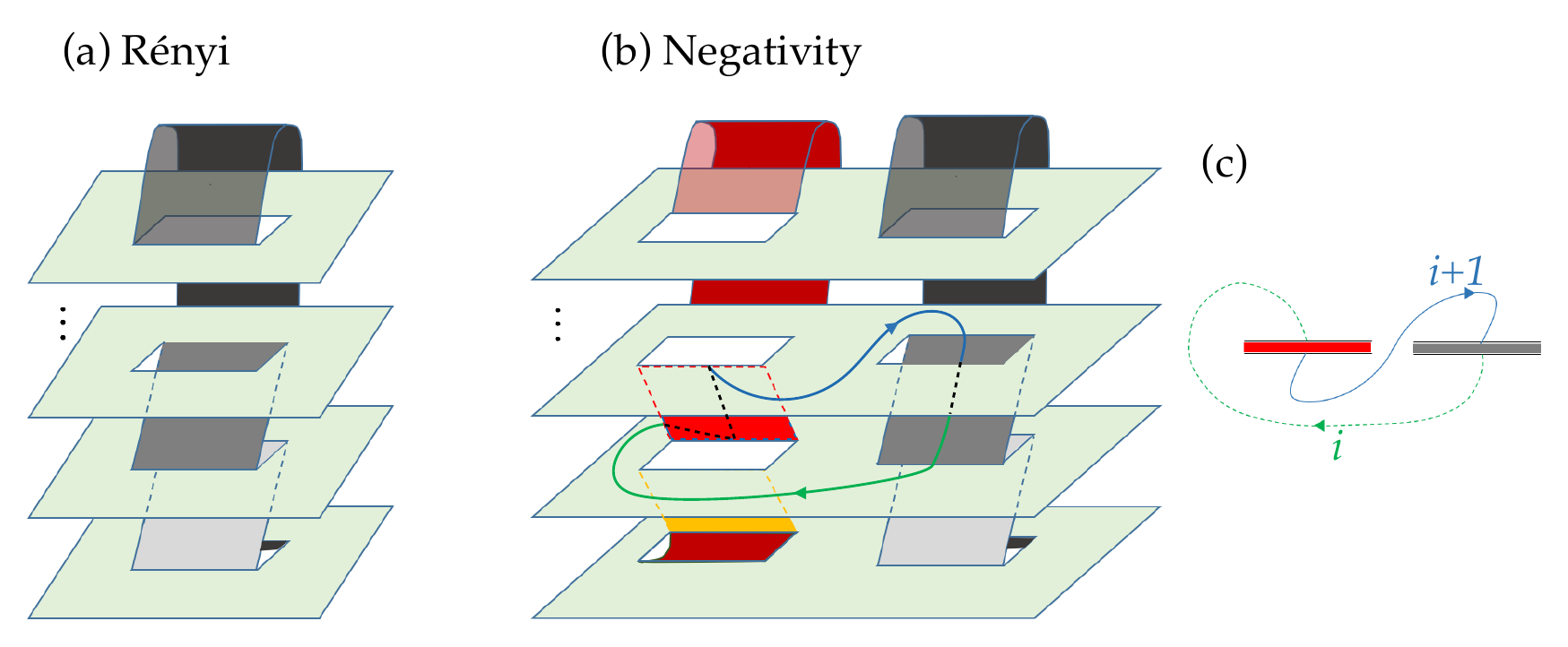}
\caption{\label{fig:spacetime} Spacetime manifolds of (a) moments of the density matrix (R\'enyi entanglement entropy) and (b) moments of the partially transposed density matrix (R\'enyi negativity). (c) A fundamental cycle of the spacetime manifold which is formed by traveling from one sheet to the next via the two intervals. An example path is shown in (b).}
\end{figure}

Before delving into the fermionic systems, let us explain the path-integral formulation~\cite{Calabrese2012,Calabrese2013} of the moments of the partially transposed density matrices $\Tr[(\rho^{\sT_1})^n]$
in bosonic systems.
 This formulation has been a particularly important milestone in generalizing the entanglement negativity to the extended systems such as spin chains and harmonic oscillator chains.
Moreover, the spacetime picture provides more intuition behind the entanglement negativity which can then be compared to other measures of entanglement such as the R\'enyi entropies (Fig.~\ref{fig:spacetime}(a)). 
According to the spacetime picture shown in Fig.~\ref{fig:spacetime}(b), the connectivity between consecutive sheets (which represent density matrices) is reversed for the transposed intervals.
This idea has made it possible to represent both direct and transposed intervals by a set of twist operators. From this, one can eventually write the moments of the partial transpose in terms of few-point correlation functions of twist operators.
The path-integral formulation has initiated several efforts in calculating the logarithmic negativity for various models including CFTs 
\cite{Calabrese2012,Calabrese2013}, massive quantum field theories~\cite{Blondeau2016} as well as finite-temperature \cite{Calabrese_Ft2015}
and out-of-equilibrium situations \cite{Coser_quench2014,Hoogeveen2015, PhysRevB.92.075109}.  

To emphasize the importance of ordering in the fermionic density matrices, let us consider the ``normal-ordered" occupation number basis as
\begin{align}
\ket{\{ n_j \}_{j \in A_1} , \{n_j\}_{j \in A_2}} 
&= (f_{m_1}^{\dag})^{n_{m_1}} \cdots (f_{m_{\ell_1}}^{\dag})^{n_{m_{\ell_1}}} \cdots (f_{m_{\ell_2}'}^{\dag})^{n_{m_{\ell_2}'}} \ket{0}
\end{align}
where $n_j$'s are occupation numbers for the subsystems $A_1$ and $A_2 $,  which contain $\ell_1$ and $\ell_2$ sites, respectively. Normal-ordering in this representation means that the fermionic degrees of freedom inside each subsystem are clustered together. 

As in Refs.~\cite{Shap_pTR,Shiozaki_antiunitary}, our guiding principle to
define the analog of partial transpose for fermionic systems is the
path-integral formulation which introduces a way to realize the partition
function of fermions on the spacetime manifold depicted in
Fig.~\ref{fig:spacetime}(b). The reversed connectivity of the first intervals in
the successive sheets can be viewed as reversing the time coordinate for the
first intervals, hence it is called partial time-reversal transformation.
In the occupation number basis, the transformation rule for the partial transpose is given by 
\begin{align}
\left( \ket{\{ n_j \}_{A_1} , \{n_j\}_{A_2}} \bra{\{ \bar n_j \}_{A_1}, \{ \bar n_j \}_{A_2} } \right)^{\sT_1} 
&=(-1)^{\phi(\{n_j\}, \{\bar n_j\})}
 \ket{\{ \bar n_j \}_{A_1} , \{n_j\}_{A_2}} \bra{\{ n_j \}_{A_1}, \{ \bar n_j \}_{A_2} },
\label{eq:app_f_21}
\end{align}
where the phase factor is
\begin{align}
\phi(\{n_j\}, \{\bar n_j\}) =& \frac{[(\tau_1+\bar{\tau}_1)\ \text{mod}\ 2]}{2} + (\tau_1+\bar{\tau}_1)(\tau_2+\bar{\tau}_2)
\end{align}
in which
$\tau_s=\sum_{j\in A_s} n_j$  and $\bar{\tau}_s=\sum_{j\in A_s} \bar{n}_j$, are the number of occupied states in the $A_s$ interval and $s=1,2$.  The fermionic partial transpose can also be motivated~\cite{Shiozaki_antiunitary} based on a consistent definition of full transpose in the Clifford algebra where we consider a Majorana representation of  the operator algebra and make use of the fact that actual Fermi systems in nature always preserve the fermion-number parity as a global symmetry~\cite{Bravyi2002}. 

It is worth mentioning the difference between the fermionic partial transpose and the bosonic partial transpose~\cite{Eisler2015} in the spacetime path integral formalism. The fermionic definition leads to a partition function, where the boundary conditions for fundamental cycles (Fig.~\ref{fig:spacetime}(c)) are fixed, i.e.~, there is only one spin structure. In contrast, the bosonic definition gives a sum of partition functions with all possible boundary conditions for cycles of type Fig.~\ref{fig:spacetime}(c), i.e.~, it is a sum over all possible spin structures~\cite{Coser2016_1,Coser2016_2}. Note that this boundary condition is independent of the global boundary conditions on each spacetime torus (see Eq.~(\ref{eq:torus_bc})).


\section{Free fermions in one dimension }
\label{sec:finT1d}

\begin{figure}
\center
\includegraphics[scale=0.65]{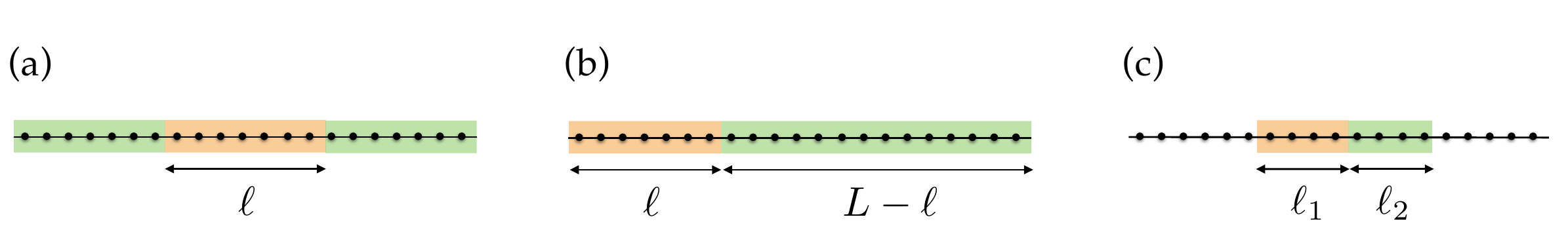}
\caption{\label{fig:chain} Various ways of partitioning the 1D system. Bipartite geometry: (a) One interval in the infinite line and (b) one interval in the semi-infinite line (open boundary condition is assumed). (c) Tripartite geometry. In all cases, \textcolor{blue}{the density matrix $\rho$ corresponds to the union of the colored regions, and} the partial transpose is applied to the orange region.}
\end{figure}

In this section, we present our results for the finite-temperature logarithmic negativity of free fermions in (1+1)d.
The Hamiltonian is given by a one dimensional lattice
with nearest neighbor hopping,
\begin{align} \label{eq:Dirac_latt}
\hat{H}= -\sum_{i} [t (f^\dag_{i+1} f_i + \text{H.c.}) + \mu f^\dag_{i} f_i ]
\end{align}
which describes a one-dimensional metal, where the Fermi surface consists of two points. The low energy theory of this model is the massless Dirac fermions in (1+1)d.
We discuss the entanglement due to three different ways of partitioning the system. Two bipartite geometries, (1) an interval of length $\ell$ embedded inside an infinite chain (Fig.~\ref{fig:chain}(a)), or (2) a semi-infinite chain (Fig.~\ref{fig:chain}(b)), and (3) a tripartite geometry with two adjacent intervals $\ell_1$ and $\ell_2$ (Fig.~\ref{fig:chain}(c)).

In Ref.~\cite{Calabrese_Ft2015}, the logarithmic negativity of a CFT of central charge $c$ was calculated at a finite temperature, when $T > 1/L$ or equivalently $L\to \infty$. It was found that the entanglement negativity of an infinite chain (Fig.~\ref{fig:chain}(a))
is given by
\begin{align} \label{eq:bi_inf}
{\cal E}(T)= \frac{c}{2} \ln \left[\frac{\beta}{\pi a} \sinh\left(\frac{\pi\ell}{\beta} \right) \right]
- \frac{\pi c\ell}{2\beta}
+ f(e^{-2\pi\ell/\beta}) + 2 \ln c_{1/2},
\end{align}
and that of a semi-infinite chain (Fig.~\ref{fig:chain}(b)) obeys the form
\begin{align} \label{eq:bi_semiinf}
{\cal E}(T)= \frac{c}{4} \ln \left[\frac{\beta}{\pi a} \sinh\left(\frac{2\pi\ell}{\beta} \right) \right]
- \frac{\pi c\ell}{2\beta}
+ f_\text{bdy} (e^{-4\pi\ell/\beta}) + 2 \ln \tilde c_{1/2},
\end{align}
where $\beta=1/T$ is the inverse temperature and $a$ is the lattice constant. In the first expression, $f(x)= \lim_{n_e\to 1} \ln[{\cal F}_{n_e}(x)]$ is a universal function which depends on the full operator content of the model and ${\cal F}_{n}(x)$ comes from the four-point function of the twist operators.
In the second expression,  $f_\text{bdy}(x)= \lim_{n_e\to 1} \ln[{\cal B}_{n_e}(x)]$ is the universal scaling function with a boundary,  which generically depends also on the boundary state of the CFT and it cannot be simply obtained from ${\cal F}_n$ in the bulk. We should notice the important difference between the above formulas where the dependence is $\sinh (2\pi\ell/\beta)$ for the semi-infinite line case instead of $\sinh (\pi\ell/\beta)$ for the case of  infinite line (\ref{eq:bi_inf}).  

In what follows, we provide a rather self-contained derivation of the logarithmic negativity associated with the fermionic partial transpose when both temperature $T$ and the system size $L$ are finite. To this end, we first review Casini's approach~\cite{Casini2005} to derive the leading order term for the R\'enyi entropy of massless Dirac fermions in (1+1) dimension~\cite{Azeyanagi,RyuTakayanagi,Ogawa2012,Herzog2013} and then apply it to compute the logarithmic negativity of the same system. We obtain analogous results to Eqs.~(\ref{eq:bi_inf}) and (\ref{eq:bi_semiinf}) for the fermionic negativity in the $L\to \infty$ limit and finite $T$.
Later on in this section, we compare the analytical results with the numerical calculation on a lattice and show that they are in agreement.

\subsection{Review of R\'enyi entanglement entropy}
\begin{figure}
\center
\includegraphics[scale=0.68]{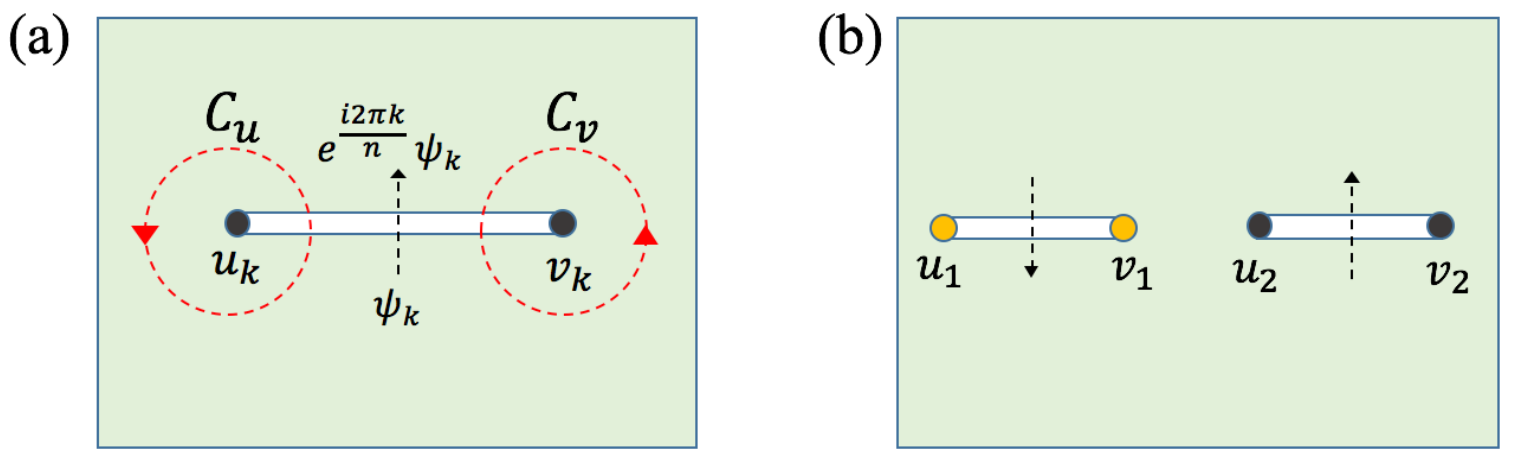}
\caption{\label{fig:casini} The twist-field picture of (a) R\'enyi entanglement entropy and (b) R\'enyi negativity, where intervals are viewed as branch cuts.}
\end{figure}

In this part, we explain how to calculate the R\'enyi entropy of massless Dirac fermions at finite temperature following Ref.~\cite{Herzog2013}.
The R\'enyi entanglement entropy of a reduced density matrix $\rho$ is defined by 
\begin{align}
S_n = \frac{1}{1-n} \ln\text{Tr} [\rho^n].
\end{align}
Using the replica approach, the above expression can be viewed in terms of $n$ replicas
\begin{align}
\Psi= (\psi_1,\cdots,\psi_n)^{\sT},
\end{align}
where $\psi_i$ and $\psi_{i+1}$ fields are identified along the interval $I_i$ which connects the $i$-th sheet to the $(i+1)$-th sheet. Each interval $I_i$ can be represented by its two endpoints $u_i$ and $v_i$.
 Alternatively, one can consider multi-valued field $\Psi$ on a single-sheet spacetime. This way, when we traverse a circuit around $u_i$ or $v_i$  (denoted as $C_{u_i}$ and $C_{v_i}$ in Fig.~\ref{fig:casini}) the  field transforms as
\begin{align}
\tilde{\psi}_i = T_{ij} \psi_j
\end{align}
where the twist matrix is given by
\begin{align} \label{eq:Tmat}
T_{ij}= -\delta_{i+1,j}\,
\end{align}
with boundary condition $n+1\equiv 1$ and $T_{n1}=1$.
For noninteracting fermions with conserved particle number,
we can diagonalize the twist matrix $T$ in Eq.~(\ref{eq:Tmat}) and rewrite the R\'enyi entropy in terms of $n$-decoupled copies,
\begin{align}
S_n = \frac{1}{1-n} \sum_{k=-\frac{n-1}{2}}^{\frac{n-1}{2}} \ln Z_k
\end{align}
where $Z_k$ is the partition function in the presence of a finite branch cut (interval) such that the phase of the fermion field is twisted as a result of passing through the interval, i.e.,~$\psi_k \to e^{i 2\pi \frac{k}{n}} \psi_k$ (see Fig.~\ref{fig:casini} (a)). The phase twist factors $e^{i 2\pi \frac{k}{n}}$ associated with $Z_k$ for $k=-\frac{n-1}{2},\cdots,\frac{n-1}{2}$ are the eigenvalues of the twist matrix. \textcolor{blue}{The $k$-mode decomposition in the replica space is generic and can be applied to interacting fermions. In the case of free fermions, we obtain $n$-decoupled theories which can be easily handled then. However, in the case of interacting fermions the four-fermion (or higher order interactions) will lead to coupling between different $k$-modes. Hence, it is not \emph{a priori} clear whether this procedure is beneficial for interacting systems.}

Following Casini~\emph{et.~al}~\cite{Casini2005},  we note that the partition function in the presence of phase twisting intervals can be formulated as a theory subject to an external gauge field.
The external gauge field is a pure gauge everywhere, except at the points $u_{i}$ and
$v_{i}$ where it is vortex-like. 
The idea is to get rid of the phase twists by performing a singular gauge transformation 
\begin{equation}
\psi_{k}(x)\to e^{i\int_{x_{0}}^{x}dx^{^{\prime }\mu }A_{\mu }^{k}(x^{^{\prime }})}\psi
_{k}\left( x\right),
\end{equation}
where $x_{0}$ is an arbitrary reference point. The new field is now single-valued everywhere.
Hence, we can absorb the boundary conditions for $C_{v_i}$ and $C_{u_i}$ circuits into an external gauge field and the resulting Lagrangian density reads
\begin{align} \label{eq:gauged_dirac}
{\cal L}_{k}=\bar{\psi}_{k}\gamma ^{\mu }\left( \partial _{\mu }+i\,A^k_{\mu}\right) \psi_{k}.
\end{align}
From the boundary conditions on $\psi_{k}$ defined over the spacetime, the gauge fields are constrained as in
\begin{align}
\oint_{C_{u_i}}dx^{\mu }A_{\mu }^{k}(x) =&-\frac{2 \pi k}{n} \,,  \label{eq:a1}
\\
\oint_{C_{v_i}}dx^{\mu }A_{\mu }^{k}(x) =&\frac{2 \pi k}{n} \,.
\label{eq:a2}
\end{align}
One subtlety in this approach is that we can add phase shifts of $2\pi
m$, with $m$ an integer, to the right hand side of the above expressions
and yet it does not change the total phase factor along the circuits. 
It turns out that this ambiguity in general leads to different representations of the partition function $Z_k$~\cite{Jin2004,Calabrese_gfc,Abanov,Ovchinnikov}. Hence, $Z_k$ can be written as a summation over all representations. The asymptotic behavior of each term in the thermodynamic limit (large (sub-)system size) is a power law $\ell^{-\alpha_m}$ and the leading order term corresponds to the smallest exponent.
For instance, the leading order term in the R\'enyi entropy of the ground state is given by the $m=0$ term.
See Appendix~\ref{sec:fisherhartwig} for more discussion on this.
 As we will see in the case of entanglement negativity,  we need to consider $m\neq 0$ for some values of $k$. 

The magnetic flux of the gauge fields satisfying Eqs.~(\ref{eq:a1}) and (\ref{eq:a2}) is given by
\begin{align}
\epsilon ^{\mu \nu }\partial _{\nu}A_{\mu }^{k}(x)=2\pi \frac{k}{n}
\sum_{i=1}^{p}\big[ \delta (x-u_{i})-\delta (x-v_{i})\big] \,.  \label{eq:bRenyi}
\end{align}
The above formula is for $p$ intervals and as we see each interval is represented by a vortex-antivortex pair
of strength $2\pi \frac{k}{n}$. The goal here is to compute the partition function as thermal expectation value in the free Dirac theory 
\begin{equation}
Z_{k}=\left\langle e^{i\int A_{\mu }^{k}j_{k}^{\mu }d^{2}x}\right\rangle \,,
\label{eq:partRenyi}
\end{equation}
where $j_{k}^{\mu }=\bar{\psi}_{k}\gamma ^{\mu } \psi_k$ is the Dirac current and $A_{\mu }^{k}$ satisfies
(\ref {eq:bRenyi}).

In order to evaluate (\ref{eq:partRenyi}), we use the
bosonization technique to express the current
$j_{k}^{\mu }$ as
\begin{equation}
j_{k}^{\mu }\to \frac{1}{2\pi}\epsilon ^{\mu \nu }\partial_{\nu}\phi_k \,,  \label{eq:corr}
\end{equation}
where $\phi_k$ is a real scalar field. For a free massless Dirac field, the
theory for the dual field $\phi_k$ is simply
\begin{align}
{\cal L}_{\phi }=\frac{1}{8\pi}\partial _{\mu }\phi \partial ^{\mu }\phi \,.
\label{eq:bosonth}
\end{align}
Therefore we have to evaluate
\begin{align}
Z_{k}=\left\langle e^{i\int A_{\mu }^{k}\frac{1}{2\pi}\epsilon ^{\mu
\nu }\partial _{\nu}\phi d^{2}x}\right\rangle = \left\langle \prod_{i=1}^{p} V_k(u_i) V_{-k}(v_i)
\right\rangle
\end{align}
where $V_k(x) =e^{-i\frac{k}{n} \phi (x)}$ is the vertex operator and the expectation values correspond to the
scalar-field theory (\ref{eq:bosonth}). Hence, the finite-temperature partition function of a Dirac fermion can be described by a compactified boson with radius $R = 2$ (Here, the self-dual radius is $R=\sqrt{2}$).
We consider the spacetime torus by identifying $z\sim z+1$ and $z\sim z+\tau$ where 
 $\tau=i\beta/L$ is the dimensionless inverse temperature. 
The fermionic theory is specified by the two boundary conditions along the two cycles of the spacetime torus
\begin{align} \label{eq:torus_bc}
\psi_k(z+1)= e^{i\pi \nu_1} \psi_k(z), \qquad
\psi_k(z+\tau)= e^{i\pi \nu_2} \psi_k(z), 
\end{align}
where $\nu_i=0$ or $1$. 
We denote the four possible boundary conditions (spin structures) by $\nu\equiv \{ \nu_1\nu_2 \}=\{ 00, 01, 11, 10\}\equiv \{1,2,3,4 \}$ cases, respectively. \textcolor{blue}{We should remind the reader that these boundary conditions are independent of the boundary conditions across the inter-spacetime cycles shown in Fig.~\ref{fig:spacetime}(c), which are fixed to be periodic.}
The correlation function of the vertex operators on the torus in sector $\nu$ is found by~\cite{yellowbook}
\begin{align}
&\braket{V_{e_1}(z_1,\bar{z}_1)\cdots V_{e_N}(z_N,\bar{z}_N)}= \prod_{i<j} \left| \frac{\partial_z \vartheta_1(0|\tau)}{\vartheta_1(z_j-z_i|\tau)} \right|^{-2e_i e_j}
\left|\frac{\vartheta_\nu(\sum_i e_i z_i|\tau )}{\vartheta_\nu(0|\tau)} \right|^2
\end{align}
in terms of the Jacobi theta functions for $R=2$ where $V_e(z,\bar{z})=e^{i e\phi(z,\bar{z})}$ is the vertex operator. Hence, we can write for the partition function in sector $\nu$
\begin{align}
Z^{(\nu)}_{k}=&\left|\frac{\prod_{i<j} \vartheta_1(u_i-u_j |\tau)  \vartheta_1(v_i-v_j|\tau) }{\prod_{i,j} \vartheta_1(u_i-v_j|\tau)} \cdot (\epsilon \partial_z \vartheta_1(0|\tau))^p \right|^{2\frac{k^2}{n^2}}  \cdot
\left| \frac{\vartheta_\nu (\frac{k}{n} \sum_i  (u_i-v_i)|\tau)}{\vartheta_\nu (0|\tau)} \right|^2,
\end{align}
where the partition function is normalized such that $Z_{k}^{(\nu)}=1$ in the absence of any branch points.
The dimensionless quantity $\epsilon=a/L$ is introduced as a UV-cutoff to resolve the coincident points since the theta function behaves as $\vartheta_1(z|\tau)\sim z$ when $z\to 0$.
As a result, the R\'enyi entropy reads
\begin{align}
S_n^{(\nu)}= S_{n,0} + S_{n,1}^{(\nu)}
\end{align}
where the first term is universal
\begin{align} \label{eq:Renyi_uni}
S_{n,0} =& -\frac{n+1}{6n} \ln {\Big|} \frac{\prod_{i<j} \vartheta_1(u_i-u_j |\tau)  \vartheta_1(v_i-v_j|\tau) }{\prod_{i,j} \vartheta_1(u_i-v_j|\tau)}  (\epsilon \partial_z \vartheta_1(0|\tau))^p  {\Big|},
\end{align}
and the second term depends on the spin structure $\nu$,
\begin{align}
S_{n,1}^{(\nu)}= \frac{2}{1-n} \sum_{k=-\frac{n-1}{2}}^{\frac{n-1}{2}}
\ln \left| \frac{\vartheta_\nu (\frac{k}{n} \sum_i  (u_i-v_i)|\tau)}{\vartheta_\nu (0|\tau)} \right|.
\end{align}
We should note that the R\'enyi entropies in the $\nu = 1$ sector are divergent,
since 
  $\vartheta_1(0|\tau)$ in the denominator of $S_{n,1}^{(\nu)}$
  is zero.
  This is related to the fermion zero mode in this sector.

Let us now look at various limits of the above result. For simplicity, we shall consider a single interval and the sector $\nu=3$ which is the usual anti-periodic boundary condition in both space and time directions.

In the low temperature limit $\tau=i\beta/L \to i\infty$, we have
\begin{align}
S_{n,0} &=  \frac{n+1}{6n} \ln \left| \frac{L}{\pi a}  \sin(\frac{\pi \ell}{L})\right| + O(e^{-2\pi/(LT)}), 
\end{align}
where we use the relation
\begin{align}
\lim_{\beta \to \infty} \vartheta_1(z|i\beta) &= 2 e^{-\pi\beta/4} \sin \pi z + O (e^{-2\pi\beta}).
\end{align} 
The second term $S_{n,1}^{(\nu)}$ is vanishing and we arrive at the usual expression for the ground state of CFT.

In the high temperature limit where $\tau=i\beta/L \to 0$, we obtain
\begin{align} \label{eq:R0highT}
S_{n,0}= \frac{n+1}{6n} \left[ -\frac{\pi \ell}{\beta}\frac{\ell}{L}+ \ln \left| \frac{\beta}{\pi a} \sinh(\frac{\pi \ell}{\beta})  \right| \right] + O(e^{-\pi LT}),
\end{align}
for the universal part and 
\begin{align} \label{eq:R1highT}
S_{n,1}^{(3)} =& \frac{(1+n)}{6n} \frac{\pi \ell^2}{\beta L}  
- \frac{2}{1-n} \sum_{j=1}^\infty \frac{(-1)^{j}}{j}\frac{1}{\sinh (\frac{\pi j}{\beta})} \left( \frac{\sinh (\frac{\pi j \ell}{\beta})}{\sinh(\frac{\pi j \ell}{n\beta})} -n\right) \, 
\end{align}
for the second part (similar results can be derived for other sectors). We should note that the first terms in  (\ref{eq:R0highT}) and (\ref{eq:R1highT}) precisely cancel each other. Here, we use the modular transformation rules for the theta functions,
\begin{align}
	\vartheta_1(z|\tau) = -(-i \tau)^{-1/2} e^{-\pi iz^2/\tau} \vartheta_1 (z/\tau | {-}1/\tau),
\end{align}
and the asymptotic form of the theta function in the small $\beta$ limit
\begin{align}
	\vartheta_1 (z/\tau | {-}1/\tau) = -2i \, e^{-\frac{\pi L}{4\beta}} \sinh (\frac{\pi z L}{\beta}) + O( e^{\frac{3\pi L}{\beta} (z-3/4)}),
\end{align}
for $0 \leq z \leq 1/2$.

\subsection{Moments of the partial transpose}

Let us recall the definitions of the moments of partial transpose for any integer powers as follows
\begin{align} \label{eq:pathTR0}
{\cal E}_n :=\left\{ \begin{array}{ll}
\ln \Tr (\rho^{\sT_1} \rho^{\sT_1\dag}  \cdots \rho^{\sT_1}\rho^{\sT_1\dag})& \ \ n\  \text{even}, \\
\ln \Tr (\rho^{\sT_1} \rho^{\sT_1\dag} \cdots \rho^{\sT_1}) &  \ \ n\  \text{odd}.
\end{array} \right.
\end{align}
The consecutive presence of $\rho^{\sT_1\dag}$ and $\rho^{\sT_1}$  leads to periodic boundary condition along the non-contractible loop between the consecutive replicas $i$ and $i+1$. 
 The zero temperature limit of the above quantity for even $n$ was previously studied in Ref.~\cite{Herzog2016}.
In order to compute the moments, we need to consider two intervals with different twist matrices $T$ and $T^R$.
The new twist matrix $T^R$ introduced for the transposed (time reversed) interval is given by
\begin{align} \label{eq:TRmat}
T^R_{ij}= \delta_{i,j+1}\ ,
\end{align}
with the boundary condition $T^R_{1n}=(-1)^{n-1}$.
Note that the overall boundary condition due to the twist matrices is identical $T^n=(T^R)^n=(-1)^{n-1}$. Fortunately, these two matrices are simultaneously diagonalizable and we can carry out a similar procedure as before by decomposing it into $n$ decoupled copies where the phase twists are different for the two intervals. We may write
\begin{align}
 {\cal E}_n= \sum_{k=-(n-1)/2}^{(n-1)/2} \ln Z_{R, k}
\end{align}
where $Z_{R,k}$ is the partition function containing  two intervals with the twist phases $e^{i2\pi (\frac{k}{n}-\varphi_n)}$ and $e^{-i2\pi \frac{k}{n}}$   where $\varphi_n=\pi$ or $\pi(\frac{n-1}{n})$ for $n$ even or odd, respectively. In the following parts, we compute the entanglement negativity for various geometries as shown in Fig.~\ref{fig:chain}.

\subsubsection{Tripartite geometry}

Here, we consider two adjacent intervals partitioned from a long chain (see Fig.~\ref{fig:chain}(c)).
Hence, the gauge field appearing in (\ref{eq:gauged_dirac}) for such configuration is given by
\begin{align}
\frac{1}{2\pi} \epsilon ^{\mu \nu }\partial _{\nu}A_{\mu }^{k}(x)=&\left(\frac{ k}{n}-\frac{\varphi_n}{2\pi} \right) \delta (x-u_1) -\left(\frac{2k}{n} -\frac{\varphi_n}{2\pi}\right) \delta (x-v_1) + \frac{k}{n}\delta (x-v_2)
  \label{eq:bNeg}
\end{align}
where we place the branch points at $u_1=-r_1$, $v_1=u_2=0$, and $v_2=r_2$. Similar to the previous derivation, we find
\begin{align}
Z^{(\nu)}_{R,k}=& \left| \vartheta_1(r_1|\tau) \right|^{-2(|\frac{k}{n}|-\frac{\varphi_n}{2\pi})(|\frac{2k}{n}|-\frac{\varphi_n}{2\pi})}
\cdot \left| \vartheta_1(r_2|\tau) \right|^{-2|\frac{k}{n}|(|\frac{2k}{n}|-\frac{\varphi_n}{2\pi})} \left| \vartheta_1(r_1+r_2| \tau) \right|^{2|\frac{k}{n}|(|\frac{k}{n}|-\frac{\varphi_n}{2\pi})} \nonumber
\\ &\times
\left|\epsilon\partial_z \vartheta_1(0|\tau)\right|^{-\Delta_k} \cdot
\left| \frac{\vartheta_\nu \left( \frac{k}{n}(r_2-r_1)+\text{sgn}(k) \frac{\varphi_n}{2\pi} r_1 |\tau\right)}{\vartheta_\nu (0|\tau)} \right|^2,
\end{align}
where 
\begin{align} \label{eq:deltak}
\Delta_k =
-\frac{6k^2}{n^2}+\frac{3\varphi_n}{2\pi} \left|\frac{k}{n}\right|-\frac{\varphi_n^2}{2\pi^2} .
\end{align}
It is important to note that for $k<0$, we modified the flux at $u_1$ and $v_1$ by inserting an additional $2\pi$ and $-2\pi$ fluxes, respectively, where the scaling exponent $\alpha_m$ takes its minimum value (refer to~\ref{sec:fisherhartwig}).
 One can also see this independently in a separate derivation in terms of the Toepltiz matrix~\cite{Shap_pTR}. Putting together, the moments of negativity are given by ${\cal E}_n={\cal E}_{n,0} +{\cal E}_{n,1} $ where the universal part is 
\begin{align} \label{eq:TRresult}
{\cal E}_{n_o,0}=& -\left(\frac{n_o^2-1}{12n_o}\right) \ln {\Big|} \vartheta_1(r_1|\tau)\vartheta_1(r_2|\tau) \vartheta_1(r_1+r_2|\tau)  \cdot (\epsilon\partial_z \vartheta_1(0|\tau))^{-3}{\Big|}, \\
{\cal E}_{n_e,0}=& -\left(\frac{n_e^2-4}{12n_e}\right) \ln {\Big|} \vartheta_1(r_1|\tau)\vartheta_1(r_2|\tau) (\epsilon\partial_z \vartheta_1(0|\tau))^{-2}{\Big|} \nonumber \\ &-\left(\frac{n_e^2+2}{12n_e}\right) \ln {\Big|} \vartheta_1(r_1+r_2|\tau)(\epsilon\partial_z \vartheta_1(0|\tau))^{-1} {\Big|} ,
\end{align} 
for $n$  odd or even and the spin structure dependent term is
\begin{align} \label{eq:Negspin}
{\cal E}_{n,1}^{(\nu)} = 2  \sum_{k=-\frac{n-1}{2}}^{\frac{n-1}{2}}  \ln \left| \frac{\vartheta_\nu \left( \frac{k}{n}(r_2-r_1)+\text{sgn}(k) \frac{\varphi_n}{2\pi} r_1 |\tau\right)}{\vartheta_\nu (0|\tau)} \right|.
\end{align}
In the above expressions, $r_i=\ell_i/L$ are the dimensionless lengths. 
Hence, the logarithmic negativity is given by ${\cal E}^{(\nu)}= {\cal E}_0+{\cal E}_1^{(\nu)}$ where
\begin{align}
 {\cal E}_0=\lim_{n_e\to 1}  {\cal E}_{n_e,0} = \frac{1}{4} \ln \left| \frac{\vartheta_1(r_1|\tau)\vartheta_1(r_2|\tau)}{\vartheta_1(r_1+r_2|\tau)} (\epsilon\partial_z \vartheta_1(0|\tau))^{-1} \right|  
\end{align}
is the universal part. There is no closed-form expression for the non-universal part ${\cal E}_1^{(\nu)}=\lim_{n_e\to 1} {\cal E}_{n_e,1}^{(\nu)}$ in a generic case other than when $\ell_1=\ell_2$. In the case of intervals with equal lengths the negativity is simplified into
\begin{align} \label{eq:Neg_tri}
{\cal E}^{(\nu)} = \frac{1}{4} \ln \left| \frac{\vartheta_1(r|\tau)^2}{\vartheta_1(2r|\tau)} (\epsilon\partial_z \vartheta_1(0|\tau))^{-1} \right|  +
 2 \ln \left| \frac{\vartheta_\nu(\frac{r}{2}|\tau)}{\vartheta_\nu(0|\tau)} \right|.
\end{align}
where $r=\ell/L$.

Let us now examine the limiting behaviors. In the low temperature limit where $\tau=i\beta/L \to \infty$, we get
\begin{align}
{\cal E}_0= \frac{1}{4} \ln \left| \frac{L}{\pi a}  \frac{\sin(\frac{\pi \ell_1}{L})\sin(\frac{\pi \ell_2}{L})}{\sin(\frac{\pi (\ell_1+\ell_2)}{L})} \right| + O(e^{-2\pi/(LT)}),
\end{align}
while in the high temperature limit where $\tau=i\beta/L \to 0$, we obtain
\begin{align}
{\cal E}_0= \frac{1}{4} \left[ \frac{2\pi \ell_1\ell_2}{\beta L}+ \ln \left| \frac{\beta}{\pi a} \frac{\sinh(\frac{\pi \ell_1}{\beta})\sinh(\frac{\pi \ell_2}{\beta})}{\sinh(\frac{\pi (\ell_1+\ell_2)}{\beta})}  \right| \right] + O(e^{-\pi LT}).
\end{align}
As we will see below, the the first term in the above expression is cancelled by the contribution from ${\cal E}_1$.
In the following, we evaluate the spin structure dependent term~(\ref{eq:Negspin}) for the $\nu=3$ sector and even $n=n_e$. Similar expressions can be derived for other sectors.
So, the low temperature limit of (\ref{eq:Negspin}) is found by
\begin{align}
 {\cal E}_{n_e,1}^{(3)} &= 2 \sum_{j=1}^\infty \frac{(-1)^{j+1}}{j\sinh (\pi \beta j)} \left( \frac{\sin(\pi j r_2 )- \sin (\pi jr_1 )}{\sin \left( \pi j(r_2-r_1)/{n_e} \right)} - {n_e} \right).
\end{align}
To the leading order, the above expression contributes to the negativity as  
\begin{align}
{\cal E}_{1}^{(3)}=\lim_{n_e\to 1}  {\cal E}_{n_e,1}^{(3)}=4 e^{-\pi/(LT)} \left(\frac{\cos(\pi (r_2+r_1)/2)}{\cos(\pi (r_2-r_1)/2)} -1 \right).
\end{align}
This is reminiscent of the universal thermal corrections found for the R\'enyi entropies~\cite{Herzog2013}.
The high temperature limit is determined by
\begin{align} 
{\cal E}_{n_e,1}^{(3)} =& -\frac{\pi}{2\beta L}\left[\left(\frac{n_e^2-1}{3n_e}\right) (\ell_2-\ell_1)^2+{n_e} \ell_1 (\ell_2-\ell_1)+ n_e\ell_1^2 \right] \nonumber \\
&-2 \sum_{j=1}^\infty \frac{(-1)^{j}}{j}\frac{1}{\sinh (\frac{\pi j}{\beta})} 
 \left(\frac{\sinh\left[ (\varphi_{n_e}\ell_1+\pi (\ell_2-\ell_1) )\frac{j}{\beta}\right]-\sinh (\varphi_{n_e}\ell_1 j/\beta)}{\sinh\left(\frac{\pi (\ell_2-\ell_1)j}{{n_e}\beta}\right)} -{n_e}\right) \, 
\end{align}
and hence, we get
\begin{align} \label{eq:E1highT}
{\cal E}_{1}^{(3)} &= -\frac{\pi \ell_1\ell_2}{2\beta L}.
\end{align}
To sum up, we have shown that the logarithmic negativity of two adjacent intervals are given by
\begin{align}
{\cal E} (LT\ll 1) &= \frac{1}{4} \ln \left| \frac{L}{\pi a}  \frac{\sin(\frac{\pi \ell_1}{L})\sin(\frac{\pi \ell_2}{L})}{\sin(\frac{\pi (\ell_1+\ell_2)}{L})} \right| + O(e^{-2\pi/(LT)}), \\
{\cal E}(LT\gg 1)&= \frac{1}{4} \ln \left| \frac{\beta}{\pi a} \frac{\sinh(\frac{\pi \ell_1}{\beta})\sinh(\frac{\pi \ell_2}{\beta})}{\sinh(\frac{\pi (\ell_1+\ell_2)}{\beta})}  \right| + O(e^{-\pi LT}),
\end{align}
in the low and high temperature regimes, which is identical to the bosonic results obtained for Harmonic chains~\cite{Eisler_Neq}.
The logarithmic negativity of two adjacent intervals with equal lengths is simplified further into
\begin{align} \label{eq:tri_universal}
{\cal E}= \frac{1}{4} \ln \left[\frac{\beta}{\pi a} \tanh\left(\frac{\pi\ell}{\beta} \right) \right] + O(e^{-\pi LT}).
\end{align}
 We should note that unlike the bipartite case in Eqs.~(\ref{eq:bi_inf}) and (\ref{eq:bi_semiinf}), there is no $f(x)$ term in the tripartite case. In the language of Ref.~\cite{Calabrese_Ft2015}, this is because we sew together only parts of spacetime sheets and taking higher powers of twist fields $T$ and $T^R$ does not create a new manifold.


\begin{figure}
\center
\includegraphics[scale=0.65]{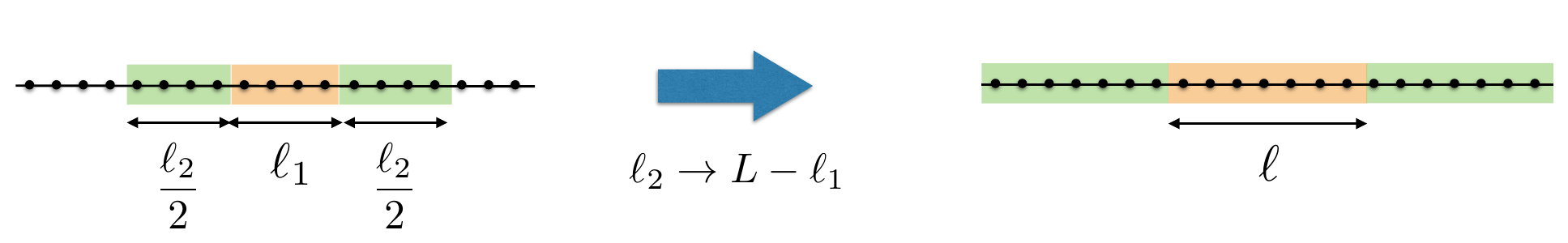}
\caption{\label{fig:bi_from_tri} The bipartite geometry (right) of a single interval on a chain of length $L$ which can be derived from the limit $\ell_2\to L-\ell_1$ of the tripartite geometry (left).}
\end{figure}

\subsubsection{Bipartite geometry}
In order to evaluate the bipartite negativity of a single interval (Fig.~\ref{fig:chain}(a)), we start by considering a tripartite geometry where an interval  of length $\ell_1$ is symmetrically embedded inside another interval of length $\ell_2$ shown in Fig.~\ref{fig:bi_from_tri}(left). Eventually, we take the limit $\ell_2\to L-\ell_1$ in our calculations, where $L$ is the total length of the chain (see Fig.~\ref{fig:bi_from_tri}).

The tripartite configuration (Fig.~\ref{fig:bi_from_tri}(left)) implies four branch points at $u_1=-r_2/2,\ v_1=0,\ u_2=r_1,$ and $v_2=r_2/2+r_1$ such that the gauge field in (\ref{eq:gauged_dirac}) must obey
\begin{align}
\frac{1}{2\pi} \epsilon ^{\mu \nu }\partial _{\nu}A_{\mu }^{k}(x)=& \left(\frac{2k}{n}-\frac{\varphi_n}{2\pi}\right) (\delta (x-v_1)-\delta (x-u_2)) + \frac{k}{n} (\delta (x-v_2) -\delta(x-u_1)).
  \label{eq:triNeg}
\end{align}
Therefore, we can find $k$-th term in the moments of partial transpose as
\begin{align}
Z^{(\nu)}_{R,k}=& \left| \vartheta_1(r_1|\tau) \right|^{-2(|\frac{2k}{n}|-\frac{\varphi_n}{2\pi})^2}\
\left| \frac{\vartheta_1(\frac{r_2}{2}|\tau)}{ \vartheta_1(\frac{r_2}{2}+r_1|\tau)} \right|^{-4|\frac{k}{n}|(|\frac{2k}{n}|-\frac{\varphi_n}{2\pi})} \cdot \left| \vartheta_1(r_1+r_2| \tau) \right|^{-2\frac{k^2}{n^2}}\
 \nonumber \\
&\times \left|\epsilon\partial_z \vartheta_1(0|\tau)\right|^{-\Delta_k} \cdot
\left| \frac{\vartheta_\nu \left( \frac{k}{n}(r_2-r_1)+\text{sgn}(k) \frac{\varphi_n}{2\pi} r_1 |\tau\right)}{\vartheta_\nu (0|\tau)} \right|^2,
\end{align}
where $\Delta_k$ is 
\begin{align}
\Delta_k =
-10\frac{k^2}{n^2}+\frac{8\varphi_n}{2\pi} \left|\frac{k}{n}\right|-\frac{\varphi_n^2}{2\pi^2} .
\end{align}
This leads to
\begin{align} \label{eq:triTR}
{\cal E}_{n_o,0}=& -\left(\frac{n_o^2-1}{6n_o}\right)  \ln {\Big|} \frac{\vartheta_1(r_1|\tau)\vartheta_1(\frac{r_2}{2}|\tau)\vartheta_1(r_1+r_2|\tau)}{\vartheta_1(\frac{r_2}{2}+r_1|\tau)} 
 (\epsilon\partial_z \vartheta_1(0|\tau))^{-2}{\Big|}, \\
{\cal E}_{n_e,0}=& -\left(\frac{n_e^2-4}{6n_e}\right) \ln {\Big|} \frac{\vartheta_1(r_1|\tau)\vartheta_1(\frac{r_2}{2}|\tau)}{\vartheta_1(\frac{r_2}{2}+r_1|\tau)} (\epsilon\partial_z \vartheta_1(0|\tau))^{-1}{\Big|}
\nonumber \\
&-\left(\frac{n_e^2-1}{6n_e}\right) \ln {\Big|} \vartheta_1(r_1+r_2|\tau)(\epsilon\partial_z \vartheta_1(0|\tau))^{-1} {\Big|} ,
\end{align} 
for the universal part and 
\begin{align}
{\cal E}_{n,1}^{(\nu)}=2\sum_{k=-\frac{n-1}{2}}^{\frac{n-1}{2}} \ln\left| \frac{\vartheta_\nu \left( \frac{k}{n}(r_2-r_1)+\text{sgn}(k) \frac{\varphi_n}{2\pi} r_1 |\tau\right)}{\vartheta_\nu (0|\tau)} \right|,
\end{align}
for the spin structure dependent term. Taking the replica limit, the universal part of the logarithmic negativity reads
\begin{align}
{\cal E}_{0}&= \frac{1}{2} \ln {\Big|} \frac{\vartheta_1(r_1|\tau)\vartheta_1(\frac{r_2}{2}|\tau)}{\vartheta_1(\frac{r_2}{2}+r_1|\tau)} (\epsilon\partial_z \vartheta_1(0|\tau))^{-1}{\Big|}.
\end{align} 
As mentioned earlier, for the bipartite geometry (Fig.~\ref{fig:chain}(a)), we need to take $r_2=1-r_1$, which gives
\begin{align}
{\cal E}_{0}&= \frac{1}{2} \ln \left| \frac{\vartheta_1(r_1|\tau)\vartheta_1(\frac{1-r_1}{2}|\tau)}{\vartheta_1(\frac{1+r_1}{2}|\tau)} (\epsilon\partial_z \vartheta_1(0|\tau))^{-1}\right| \nonumber \\
&=\frac{1}{2} \ln \left| \frac{\vartheta_1(r_1|\tau)}{ \epsilon\partial_z \vartheta_1(0|\tau)}\right|,
\label{eq:E0_inf}
\end{align} 
where
in the second line, we use the properties of theta functions to further simplify the original expression.
The low temperature limit of the second term is 
\begin{align}
{\cal E}_{1}^{(3)}=\lim_{n_e\to 1}  {\cal E}_{n_e,1}^{(3)}=4 e^{-\pi/(LT)} \left(\frac{\cos(\pi (r_2+r_1)/2)}{\cos(\pi (r_2-r_1)/2)} -1 \right).
\end{align}

Let us now  derive the high temperature expansion of the bipartite entanglement negativity. Similar to the previous section, the universal part can be simplified into
\begin{align}
{\cal E}_0= \frac{1}{2} \left[- \frac{\pi \ell_1^2}{\beta L}+ \ln \left| \frac{\beta}{\pi a} {\sinh(\frac{\pi \ell_1}{\beta})} \right| \right] + O(e^{-\pi LT}).
\end{align}
The second term in the $\nu=(3)$ sector can be evaluated as follows 
\begin{align} \label{eq:E1_bi}
{\cal E}_{n_e,1}^{(3)} =& -\frac{\pi}{2\beta L}\left[\left(\frac{n_e^2-1}{3n_e}\right) (L-2\ell_1)^2+{n_e} \ell_1 (L-2\ell_1)+ n_e \ell_1^2 \right] \nonumber \\
&-2 \sum_{j=1}^\infty \frac{(-1)^{j}}{j}\frac{1}{\sinh (\frac{\pi j}{\beta})} \left(\frac{\sinh\left[ (\varphi_{n_e}\ell_1+\pi (L-2\ell_1) )\frac{j}{\beta}\right]-\sinh (\varphi_{n_e}\ell_1 j/\beta)}{\sinh\left(\frac{\pi (L-2\ell_1)j}{{n_e}\beta}\right)} -{n_e}\right) \, 
\end{align}
which simplifies into
\begin{align}
{\cal E}_{1}^{(3)}=\frac{\pi \ell_1}{2\beta L} \left(\ell_1 -L \right) + O(e^{-\pi LT}).
\end{align}
Therefore, we find that
\begin{align} \label{eq:highT_inf}
{\cal E} (LT \gg 1)= \frac{1}{2} \left[  \ln \left| \frac{\beta}{\pi a} \sinh(\frac{\pi \ell_1}{\beta})  \right|- \frac{\pi \ell_1}{\beta} \right] + O(e^{-\pi LT}),
\end{align}
which agrees with Eq.~(\ref{eq:bi_inf}) derived in Ref.~\cite{Calabrese_Ft2015}. \textcolor{blue}{We also observe that $f(e^{-2\pi\ell/\beta})=0$ in Eq.~(\ref{eq:bi_inf}) for free fermions, since any correction to the leading order is at least of order $e^{-\pi L T}$ which is exponentially small in the total system size and vanishes in the thermodynamic limit $L/\beta\to \infty$. This is expected since the four-point function of free fermions is simple and can be reduced into a product of two-point functions.}

From the above result (\ref{eq:E0_inf}), we can find the negativity of the semi-infinite geometry by noting that the semi-infinite geometry (Fig.~\ref{fig:chain}(b)) can be obtained from the infinite geometry (Fig.~\ref{fig:chain}(a)) by cutting the $\ell_1$ interval in half; therefore, the entanglement of each cut is equal to half of the value in Eq.~(\ref{eq:E0_inf}). This gives the entanglement of a finite interval with length $\ell_1/2$ to a semi-infinite interval. Hence, the entanglement negativity of a finite interval with length $\ell_1$ to an adjacent semi-infinite interval becomes
\begin{align}
{\cal E}_0 &=\frac{1}{4} \ln \left| \frac{\vartheta_1(2r_1|\tau)}{ \epsilon\partial_z \vartheta_1(0|\tau)}\right|,
\end{align}
 the high temperature limit of which is given by
 \begin{align} \label{eq:highT_semiinf}
{\cal E} (LT \gg 1)= \frac{1}{4} \left[  \ln \left| \frac{\beta}{\pi a} \sinh(\frac{2\pi \ell_1}{\beta})  \right|- \frac{2\pi \ell_1}{\beta} \right] + O(e^{-\pi LT}).
\end{align}
This is also in agreement with the bosonic results~(\ref{eq:bi_semiinf}) of Ref.~\cite{Calabrese_Ft2015}.


\subsection{Numerical Results}

In this part, we numerically calculate the logarithmic negativity associated with the partial transpose (\ref{eq:app_f_21}) for free fermions on a lattice and compare it with the analytical results derived above.

Let us first sketch the steps of numerically computing the partial transpose and entanglement negativity for non-interacting systems.  The following procedure works efficiently for any quadratic Hamiltonian of the form $\hat{H}= \sum_{i,j}  t_{ij} f^\dag_i f_j + \text{H.c.}$ which spans a variety of non-interacting models with a conserved total particle number. A more general procedure which does not require particle number conservation is discussed in Ref.~\cite{Shap_pTR} and is briefly reviewed in~\ref{sec:app_pTR}.
The reduced density matrix can be completely characterized by the single particle correlation function~\cite{Peschel_Eisler2009}
\begin{align} \label{eq:Cmat}
C_{ij} = \braket{f_i^\dag f_j}= \text{tr} (\rho f_i^\dag f_j),
\end{align}
for the ground state $\rho=\ket{GS}\bra{GS}$ or a thermal ensemble described by the density matrix $\rho=e^{-\beta \hat{H}}$. For a thermal state, the single-particle correlator reads 
\begin{align}
C_{ij}=\sum_n f(\epsilon_n)\ u_n^\ast(i)\cdot u_n(j),
\end{align}
where $\ket{u_n}$ are single particle eigenstates $H\ket{u_n}=\epsilon_n \ket{u_n}$, $u_n(j)=\braket{j|u_n}$ is the value of the wave function at site $j$, and $f(x)=(1+\exp(\epsilon_n/T))^{-1}$ is the Fermi-Dirac distribution. For the ground state, the Fermi-Dirac distribution  at zero temperature automatically enforces the summation to be over the occupied states. 

 Let $\Gamma=\mathbb{I}-2 C$ be the covariance matrix and consider a density matrix  $\rho$ on the system  $A = A_1 \cup A_2$ 
with a covariance matrix 
\begin{align}
\Gamma = \left( \begin{array}{cc}
{ \Gamma^{11}} & {\Gamma^{12}} \\
{\Gamma^{21}} &{\Gamma^{22}} 
\end{array} \right),
\end{align}
where $\Gamma^{11}$ and $\Gamma^{22}$ denote the reduced covariance matrices of subsystems
$A_1$ and $A_2$, respectively; while $\Gamma^{12}$ and $\Gamma^{21}$ contain 
the expectation values of mixed quadratic terms. We define the transformed matrices as
\begin{align}
\Gamma_{\pm} = \left( \begin{array}{cc}
{ -\Gamma^{11}} &\pm i {\Gamma^{12}} \\
\pm i {\Gamma^{21}} &{\Gamma^{22}} 
\end{array} \right),
\end{align}
corresponding to partial transpose of the density matrix with respect to $A_1$.  Using algebra of a product of Gaussian operators~\cite{Eisert2016}, the new single particle correlation function associated with the normalized composite density operator $\Xi=\rho^{\sT_1}\rho^{\sT_1\dag}/{\cal Z}_\Xi$ can be found by
\begin{align} \label{eq:Cx}
C_\Xi=\frac{1}{2} \left[\mathbb{I} - (\mathbb{I}+\Gamma_+\Gamma_-)^{-1}(\Gamma_++\Gamma_-) \right],
\end{align}
where the normalization factor is ${\cal Z}_\Xi=\tr(\Xi)=\tr(\rho^2)$.
The entanglement negativity is then obtained by~\cite{Eisert2016}
\begin{align}
{\cal E}=\text{tr}(\sqrt{\rho^{\sT_1}\rho^{\sT_1\dag}})&= \ln \left[{{\cal Z}_\Xi^{1/2} \text{tr}(\Xi^{1/2})}\right]
\nonumber \\
&= \ln \text{tr}(\Xi^{1/2})+ \frac{1}{2}\ln \text{tr}(\rho^2).
\end{align}
In terms of eigenvalues of correlation matrices, we can write
\begin{align}
{\cal E}= \sum_j  \ln \left[ \xi_j^{1/2} + (1-\xi_j)^{1/2} \right]+
\frac{1}{2} \sum_j  \ln \left[ \zeta_j^2 + (1-\zeta_j)^2 \right],
\end{align}
where  $\zeta_j$ and $\xi_j$ are eigenvalues of the original correlation matrix $C$ (\ref{eq:Cmat}) and the transformed matrix $C_\Xi$ (\ref{eq:Cx}), respectively.


\begin{figure}
\center
\includegraphics[scale=0.7]{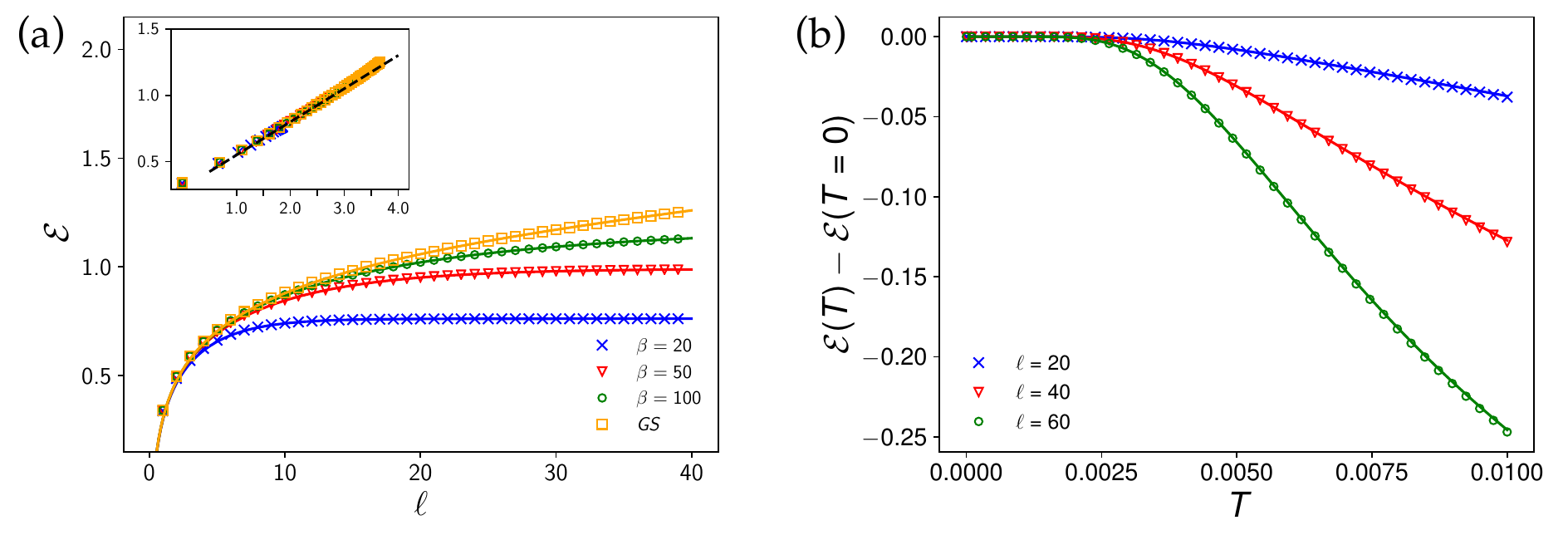}
\caption{\label{fig:tripartite}  The logarithmic negativity of the tripartite
  geometry for two adjacent intervals with equal length $\ell$ on the free
  fermion lattice model (\ref{eq:Dirac_latt}). The negativity as a function of
  (a) interval length  and (b) temperature. The solid lines are analytical
  results (\ref{eq:Neg_tri}). The inset of panel (a) shows the linear behavior
  when $\cal E$ is plotted against  $x=\ln \left[\frac{\beta}{\pi}
    \tanh\left(\frac{\pi\ell}{\beta} \right) \right]$. The dashed line in the
  inset is a reference line with the slope of $c/4=1/4$. We set $L=200$.}
\end{figure}

\begin{figure}
\center
\includegraphics[scale=0.7]{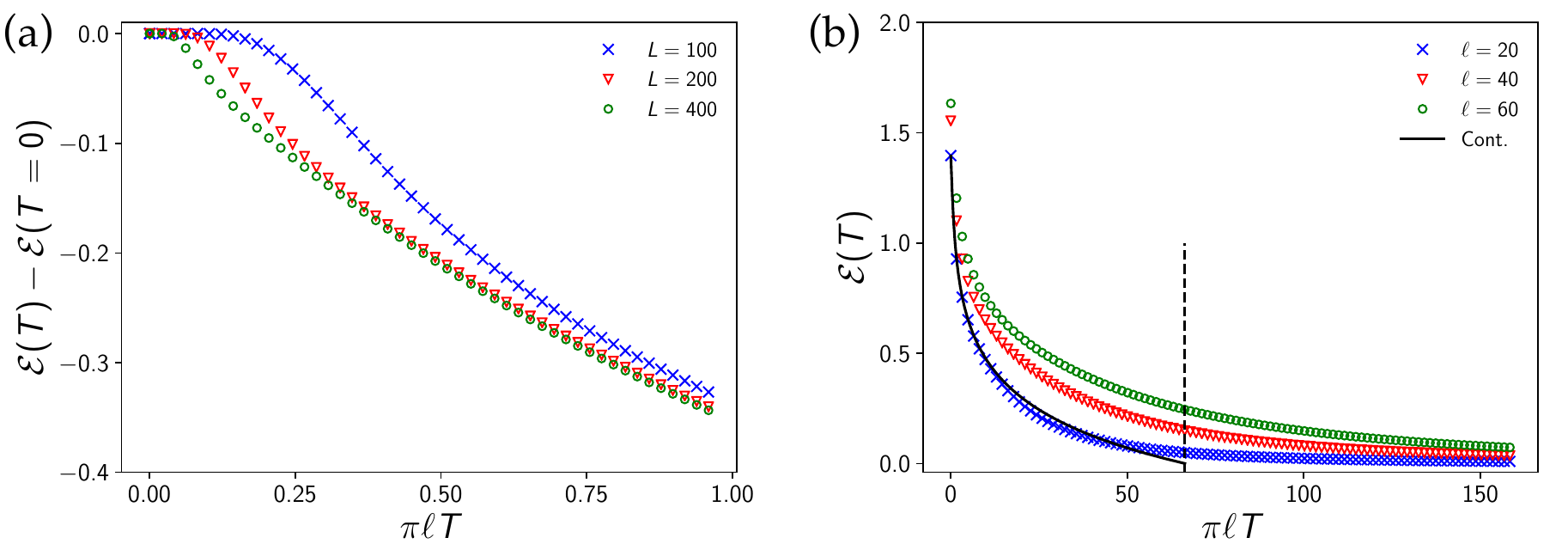}
\caption{\label{fig:bipartite_fin} The logarithmic negativity of the bipartite geometry for one interval in the semi-infinite line (Fig.~\ref{fig:chain}(b)).  (a) The subsystem size $\ell=20$ is fixed, while $T$ is varied for different total system sizes $L$. The initial plateau is due to finite length of chain. (b) The total system size is fixed, while $T$ is varied for different subsystem sizes $\ell$. Absence of sudden death in the fermionic negativity is evident. The dashed vertical line indicates the point at which the continuum limit expression~(\ref{eq:bi_semiinf}) starts to become negative. }
\end{figure}

For numerical purposes, we choose $t=1$ and set $\mu=0$ in the lattice Hamiltonian~(\ref{eq:Dirac_latt}).
In Fig.~\ref{fig:tripartite}, we compute the logarithmic negativity for two adjacent intervals (Fig.~\ref{fig:chain}(c)). The analytical results (solid lines) and lattice calculations (points) match over a  range of temperatures and interval lengths. 
As we see in Fig.~\ref{fig:tripartite}(a), the negativity saturates, i.e., obeys an area law, at any finite temperature once $\ell \gg 1/T$. This means that at finite temperatures the quantum coherence can only be maintained for length scales of order $\beta=1/T$, beyond which thermal fluctuations completely wash it out. Another interesting point is the collapse of high-temperature negativity for various values of $\ell$ and $T$ onto the universal curve (\ref{eq:tri_universal}) as shown by the inset of Fig.~\ref{fig:tripartite}(a).
We also observe that the negativity curves (e.g., Fig.~\ref{fig:tripartite}(b)) generically start off with a plateau at low temperatures. This corresponds to the limit $T\ll 1/L$  when the temperature is less than the finite-size gap ($\sim 1/L$) in the energy spectrum and the system behaves as if it is at the zero temperature.  Figure~\ref{fig:bipartite_fin}(a) illustrates this finite-size effect, where the negativity remains a plateau until some temperature, and then it falls off. The onset of decay (length of the finite-size plateau) is decreased as we make the system larger.

Next, we look at the high temperature limit in Fig.~\ref{fig:bipartite_fin}(b), where we observe that the fermionic logarithmic negativity decays to zero smoothly without any sudden death, as opposed to the bosonic case~\cite{Calabrese_Ft2015}. We believe that this is related to the fermion-number parity constraint on the fermionic density matrices~\cite{Shap_sep}. 
In~\ref{app:suddendeath}, we explicitly show in a system of two fermionic sites how this constraint limits the form of density matrix and modifies the notion of separability, which ultimately lead to the absence of sudden death.
Further, it is evident that for high temperatures and at a fixed $\ell$, the continuum limit Eq.~(\ref{eq:highT_semiinf}) is valid only until we reach the lattice scale (i.e., as long as $\beta\gg a$).

\begin{figure}
\center
\includegraphics[scale=0.7]{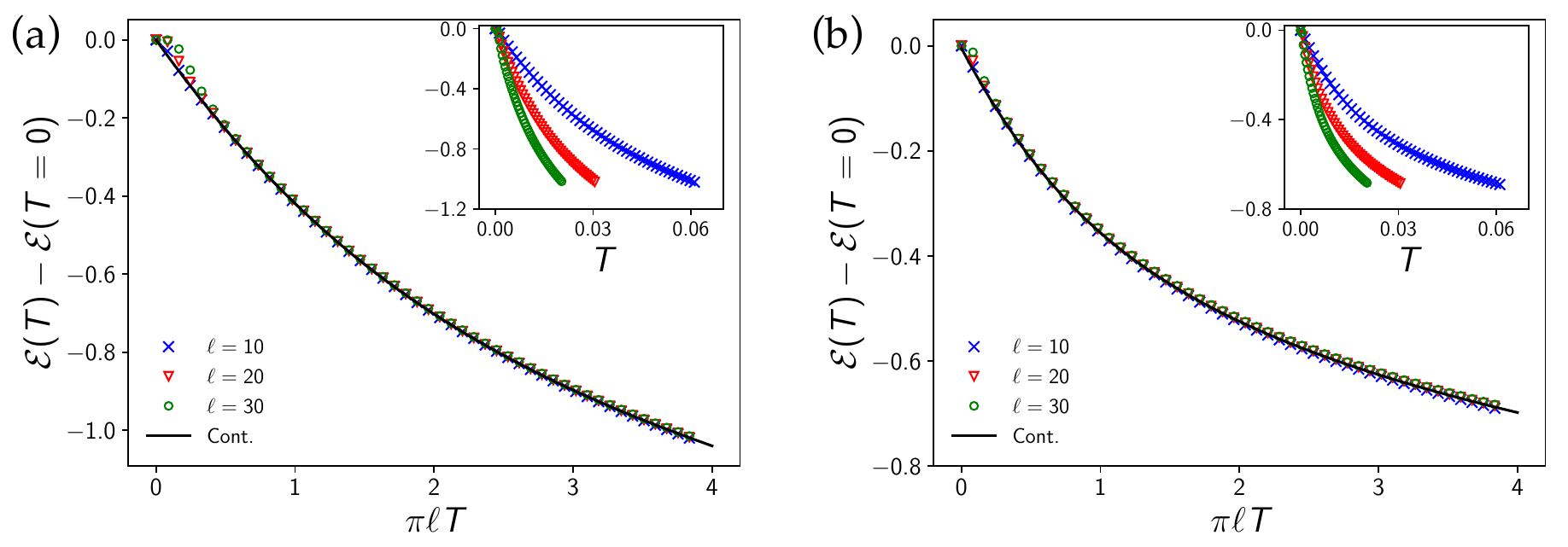}
\caption{\label{fig:bipartite} The collapse of logarithmic negativity of the bipartite geometry on universal curves.  (a) One interval in the infinite line (Fig.~\ref{fig:chain}(a)).  (b) One interval in the semi-infinite line (Fig.~\ref{fig:chain}(b)).
In each panel, the inset shows the $\cal E$ against temperature (without scaling). This is to be compared with the universal curve when the horizontal axis is normalized as $\pi\ell T$.  Here, the numerics are done for the free fermion lattice model with $L=400$.}
\end{figure}

An important implication of Eqs.~(\ref{eq:bi_inf}) and (\ref{eq:bi_semiinf}) for the logarithmic negativity of the bipartite geometries (c.f., Fig.~\ref{fig:chain}(a) and (b)) is that the entanglement difference ${\cal E}(T)-{\cal E}(0)$ is a universal function of $\pi\ell T$. 
We verify this behavior in Fig.~\ref{fig:bipartite} by looking at various interval sizes and showing that they all collapse on the same curve. For reference, we also plotted the analytical expressions (\ref{eq:highT_inf}) and (\ref{eq:highT_semiinf}). 

\section{Free fermions in higher dimensions}
\label{sec:finT2d}

In this section, we would like to extend our result for the finite-temperature entanglement negativity of one-dimensional metals to higher dimensions. 
Our idea is motivated by similar results discussed by Swingle~\cite{Swingle2010,Swingle2010_Renyi} for generalizing entanglement entropies to higher dimensional Fermi surfaces. 

The general result for the zero-temperature is as follows: the R\'enyi entanglement entropy of a subregion of size $\ell$ for a $(d+1)$ metallic system with a codimension one Fermi surface is \cite{Klich2006},
\begin{align} \label{eq:NdRenyi}
S_n(\ell)= C_d(\mu)\ \left(\frac{n+1}{6n}\right) \ell^{d-1} \ln \ell,
\end{align}
where
\begin{align} \label{eq:NdRenyi_slope}
C_d(\mu)=\frac{1}{4(2\pi)^{d-1}} \int_{\partial \Omega} \int_{\partial \Gamma(\mu)} \, dS_k dS_x |n_x \cdot n_k|,
\end{align}
$\Omega$ is the volume of the subregion normalized to one,
$\Gamma(\mu)$ is the volume enclosed by the Fermi surface, and the
integration is carried out over the surface of both domains.

In particular, the entanglement entropy of a two dimensional metal reads as
\begin{align} \label{eq:2dRenyi}
S_n(\ell)=\left(\frac{n+1}{6n}\right) C_2 (\mu)\cdot \ell \ln \ell.
\end{align}
The filled Fermi surface of a two dimensional metal may be viewed as a collection of one dimensional gapless modes~\cite{Benfatto,Shankar,Polchinski} and the entanglement can be understood as a sum of 1D segments ($\ell$ of them) each of which contributes $(n+1)/6n\cdot \ln (\ell)$ up to a geometrical coefficient (\ref{eq:NdRenyi_slope}).
The above formula was shown to be in a remarkable agreement with numerical simulations of various microscopic lattice models~\cite{Li2006,Barthel2006}. As we have seen for a one-dimensional metal, the finite temperature R\'enyi entropy has the same form as the zero temperature entropy provided that we replace $\ln (\ell)$ by $\ln [(\beta/\pi) \sinh(\pi\ell/\beta)] $. Hence,  we can follow the same lines of argument as those we use for the zero temperature to deduce that the R\'enyi entropy of a two-dimensional metal should obey the following form~\cite{Swingle2010,Swingle2010_Renyi},
\begin{align} \label{eq:2dRenyi_T}
S_n(\ell,T)=\left(\frac{n+1}{6n}\right) C_2 (\mu)\cdot \ell \ln \left| \frac{\beta}{\pi a} \sinh\left(\frac{\pi\ell}{\beta}\right)\right|.
\end{align}

Let us now consider the bipartite logarithmic negativity.
By a similar reasoning, we expect that the two dimensional negativity should obey the same form in terms of the one dimensional negativity. So, for finite temperature negativity we can write
\begin{align} \label{eq:2dNeg_T}
{\cal E}(\ell,T) =C_2 (\mu)\cdot \frac{\ell}{2} \left[ \ln{\left(\frac{\beta }{\pi a} \sinh{\left(\frac{\pi \ell}{\beta }\right)}\right)} - \frac{\pi\ell}{\beta } \right].
\end{align}
We should note that the bipartite logarithmic negativity is equal to the $1/2$-R\'enyi entropy at zero temperature.
However, there is an important difference between the logarithmic negativity and the R\'enyi entropy. 
Entanglement negativity has an extra term linear in $\ell$ inside the parenthesis compared to the $1/2$-R\'enyi entropy and this term exactly cancels the volume law term in the high temperature limit, i.e., the entanglement negativity obeys an area law ${\cal E} (\ell T \gg 1) \propto \ell \ln (\beta/\pi a)$
 while R\'enyi entropy grows as a volume law.

\begin{figure}
\center
\includegraphics[scale=0.85]{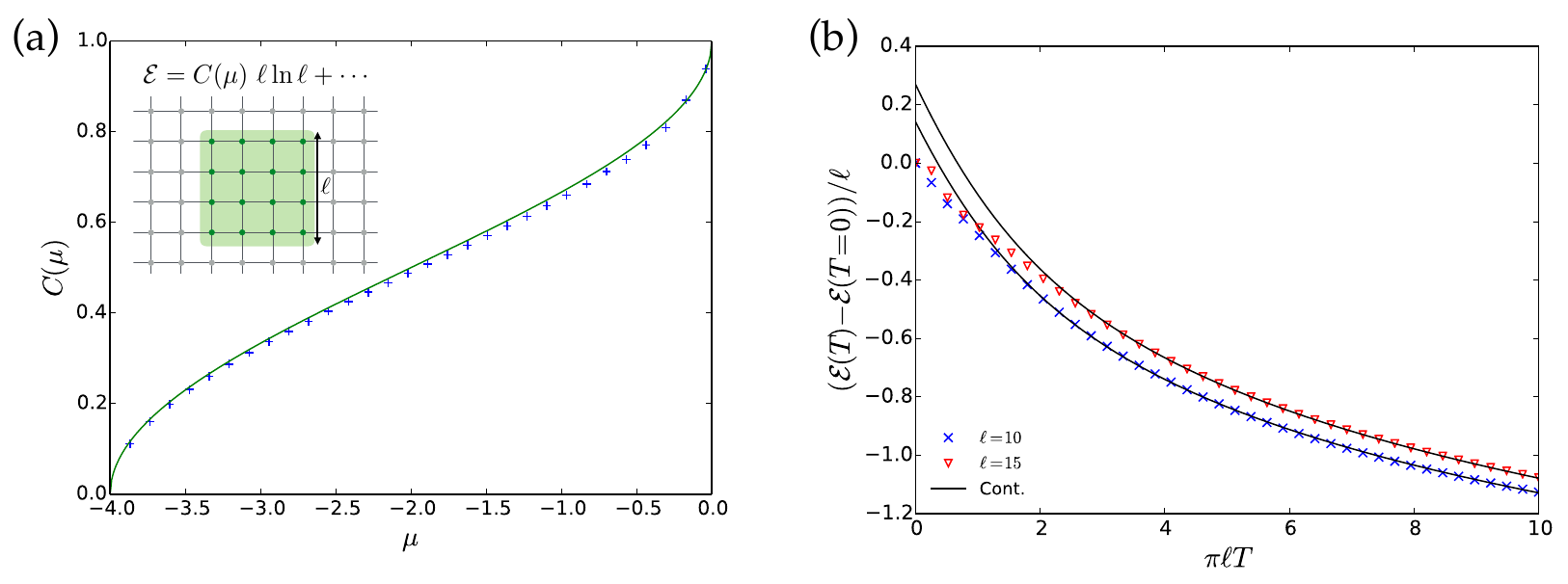}
\caption{\label{fig:2d_lattice}  The logarithmic negativity of the bipartite
  geometry on the 2d square lattice. (a) The slope of zero-temperature
  negativity as a function of chemical potential. The solid line is the
  continuum expression~(\ref{eq:2dRenyi_slope}). Here, we carry out numerics for
  the infinite size system and the slope is found for ${\cal E}$ vs. $\ell$ over
  the range $40 \leq \ell \leq 80$. (b) Finite temperature negativity as a
  function of rescaled (dimensionless) temperature. The solid lines are
  Eq.~(\ref{eq:2dNeg_T}) which are shifted vertically to fit the low-temperature
  value. The system size is $30\times 30$.
}
\end{figure}

\subsection{Numerical results}

A canonical model to benchmark two-dimensional entropies is a simple tight-binding model on a 2d square lattice described by the Hamiltonian,
\begin{align}
\hat H= - \sum_{\xv,\ev_i} [f_{\xv+\ev_i}^\dag f_{\xv}+f_{\xv}^\dag f_{\xv+\ev_i}] -\mu \sum_\xv f_\xv^\dag f_\xv.
\end{align}
The zero-temperature R\`enyi negativities of this model  for various tripartite geometries have been  studied in~\cite{PhysRevB.93.115148}.
The geometric coefficient in the entanglement entropy (\ref{eq:NdRenyi_slope}) of a square segment is found to be 
\begin{align} \label{eq:2dRenyi_slope}
C_2 (\mu)= \frac{2}{\pi} \left[ \pi- \cos^{-1}(\frac{\mu}{2}+1)\right].
\end{align}
 The bipartite entanglement negativity of a square subregion is computed in Fig.~\ref{fig:2d_lattice}. As mentioned, the zero-temperature negativity is the same as $1/2$-R\'enyi entropy and we confirm in Fig.~\ref{fig:2d_lattice}(a) that it obeys Eq.~(\ref{eq:2dRenyi}) over a wide range of chemical potential. The agreement between the numerically computed negativity at finite temperature and analytical results in the continuum limit~(\ref{eq:2dNeg_T}) is shown in Fig.~\ref{fig:2d_lattice}(b).


\section{Conclusions}
\label{sec:conclusions}

In summary, we study the entanglement negativity of free fermions with a Fermi surface of codimension one. We observe that there are two temperature regimes which are defined by the ratio $L_\beta/L$ of the length scale associated with the thermal fluctuations $L_\beta=\hbar v_F/k_B T$ to the system size $L$.
We find that the leading order term in the negativity (irrespective of dimensionality) consists of two kinds of contributions: a logarithmic term and a linear term (volume law) in system size. The logarithmic term is simplified to the usual logarithmic violation of area law at zero (and low) temperature, while it becomes a volume law at high temperatures (when $L_\beta\ll L$). 
It is interesting that these two contributions cancel each other at high temperatures which implies that the entanglement negativity asymptotically obeys an area law ${\cal E}\propto L^{d-1} \ln L_\beta$.
The area law can be understood in a qualitative picture where the entanglement comes exclusively from a strip of  width $L_\beta$ along the entanglement cut.
Overall, the decaying behavior of the negativity coincides with the fact that although finite temperature states of quantum systems are highly correlated, these correlations are mostly due to thermal fluctuations and do not include any amount of quantum entanglement.
In other words, it indicates the very characteristic of the logarithmic negativity as a measure of quantum entanglement.
Moreover, this property of the negativity is in stark contrast with the entanglement entropies such the von Neumann and R\'enyi entropies where there is only the logarithmic contribution which eventually turns into a volume law at high temperatures.

The above observation suggests that the entanglement negativity could shed light on the quantum-to-classical crossover~\cite{SwingleSenthil} as the temperature is increased. 
The cross-over temperature can then be studied in various phases of a given phase diagram away from the critical point.
For example, one direction could be to investigate the finite-temperature states of Hamiltonians which have symmetry protected topological phases or topological order as their ground states. The question is to what extent the entanglement properties of such phases survive at non-zero temperatures.
Along the same line, it would also be interesting to use the current framework and study massive quantum field theories~\cite{Blondeau2016} at finite temperatures. Out-of-equilibrium dynamics \cite{Eisler_Neq,Coser_quench2014,Hoogeveen2015, PhysRevB.92.075109} of fermions  is another interesting avenue for research.

Throughout all numerical simulations of fermionic systems in this paper, we have not experienced any sudden death in the entanglement negativity. For a system of two fermionic site, we provide an explanation based on the modified separability criterion for fermionic density matrices. We expect (but not argue) that the absence of sudden death carries through all fermionic systems with generic hopping terms~\cite{Shap_sep}. 
A rigorous proof of the absence of sudden death in the entanglement negativity of a generic extended fermionic system with many sites or a counter example in which sudden death is observed could in principle be very enlightening.

\section*{Acknowledgments} 
The authors would like to acknowledge fruitful discussions with E. Tonni, Z.~Zimbor\'as, K.~Shiozaki, X.~Wen, T.H.~Hsieh,  P.~Ruggiero, P.-Y.~Chang D.~Chowdhury, J.~Kudler-Flam, and M.-T.~Tan.
H.S. would like to thank Pasquale Calabrese for insightful discussions and constructive comments on the results, and for pointing out the previous literature about the generalized Fisher-Hartwig conjecture.
This work was supported in part by the National Science Foundation
under Grant No.\ DMR-1455296,
and 
under Grant No.\ NSF PHY-1748958.
We are grateful to the KITP Program
{\it  Quantum Physics of Information}
(Sep 18 - Dec 15, 2017), where some part of the work was performed.
We thank the Galileo Galilei Institute for Theoretical Physics for the hospitality and the INFN for partial support during the completion of this work.
H.S. acknowledges the support from the ACRI fellowship (Italy) and the KITP graduate fellowship program. 

\appendix

\section{Full expansion of R\'enyi entropy and negativity}
\label{sec:fisherhartwig}

As mentioned in the main text, the flux vortices in the decomposition of the R\'enyi entropies (or negativities) into decoupled spacetime sheets are defined modulo $2\pi$.
 Here, we explain how to find the leading order term out of all possible flux vorticities. 

Let us discuss this expansion for a generic case.
Suppose we have a partition function on a multi-sheet geometry (for either R\'enyi entropy or negativity) ${\cal S}_n$. We consider a decomposition of this quantity after diagonalizing the twist matrices as in
\begin{align}
{\cal S}_n= \sum_{k} \ln Z_k,
\end{align}
where $Z_k$ is the partition function in the presence of $2p$ flux vortices corresponding to the two ends of $p$ intervals defined between a pair of points $u_{2i-1}$ and $u_{2i}$,
\begin{equation}
Z_{k}=\left\langle e^{i\int A_{k,\mu } j_{k}^{\mu }d^{2}x}\right\rangle \,,
\end{equation}
in which 
\begin{align}
\epsilon ^{\mu \nu }\partial _{\nu}A_{k,\mu }(x)=2\pi 
\sum_{i=1}^{2p} \nu_{k,i} \delta (x-u_{i}) \,. 
\end{align}
and $2\pi \nu_{k,i}$ is vorticity of gauge flux determined by the eigenvalues of the twist matrix. 
The vorticities satisfy the neutrality condition $\sum_i \nu_{k,i}=0$ for every $k$. We observe that an arbitrary multiple of  $2\pi$ flux can be added to each vortex without changing the resulting partition function. 
In other words, there are several representations of the partition function $Z_k$. In order to obtain the asymptotic behavior, one needs to take the sum over all the representations (i.e., flux vorticities mod $2\pi$),
\begin{align}
\widetilde{Z}_k= \sum_{\{m_i\}} Z_k^{(m)} 
\end{align}
where $\{m_i\}$ is a set of integers and
\begin{equation}
Z^{(m)}_{k}=\left\langle e^{i\int A_{k,\mu }^{(m)}j_{k}^{\mu }d^{2}x}\right\rangle \,,
\end{equation}
is the partition function for the following fluxes, 
\begin{align}
\epsilon ^{\mu \nu }\partial _{\nu}A_{k,\mu }^{(m)}(x)=2\pi 
\sum_{i=1}^{2p} \widetilde\nu_{k,i} \delta (x-u_{i}),
\end{align}
and $\widetilde\nu_{k,i}=\nu_{k,i}+ m_i$ are shifted flux vorticities.
The neutrality condition implies that $\sum_i m_i=0$. Using the bosonization technique, we may write
\begin{align}
Z_k^{(m)} = E_{\{m_i\}} \prod_{i<j} \frac{1}{|u_i-u_j|^{-2\widetilde\nu_{k,i}\widetilde\nu_{k,j}}} 
\end{align}
in which $E_{\{m_i\}}$ is a constant depending on cutoff and microscopic details in the case of lattice models. To see the leading order term in the thermodynamic limit $\ell \to \infty$, we look at the scaling dimensions of each term,
\begin{align}
Z_k &\sim \sum_{\{m_i\}} \frac{E_{\{m_i\}} }{\ell^{-2\sum_{i<j} \widetilde\nu_{k,i}\widetilde\nu_{k,j} }}\\
&=  \sum_{\{m_i\}} \frac{E_{\{m_i\}} }{\ell^{\sum_{i} \widetilde\nu_{k,i}^2 }}
\end{align}
where $\ell$ is a length scale and we make use of the identity $-2\sum_{i<j} \widetilde\nu_{k,i}\widetilde\nu_{k,j} = \sum_{i} \widetilde\nu_{k,i}^2$.
From this expansion, the leading order term in the limit $\ell\to \infty$ is the one(s) which minimizes the quantity $\sum_{i} \widetilde\nu_{k,i}^2$, i.e.~the sum of squares. This is the same condition as what one gets from the generalized Fisher-Hartwig conjecture~\cite{BasTr,Basor1979}.
This point was also highlighted in previous studies~\cite{Calabrese_gfc,Abanov,Ovchinnikov}.



\section{Fermionic partial transpose at finite temperature}
\label{sec:app_pTR}

We consider a general form of quadratic Hamiltonians,
\begin{align}
\hat{H}= \sum_{i,j}  t_{ij} f^\dag_i f_j +\Delta_{ij} f^\dag_i f_j^\dag + \text{H.c.}
\end{align}
The reduced density matrix of such Hamiltonians can also be recast in a quadratic form 
\begin{align}  \label{eq:red_den}
\rho =\frac{e^{-\hat{H}_E}}{\cal Z}
\end{align}
where the entanglement Hamiltonian is $\hat{H}_E= \sum_{i,j} {h}^1_{ij} f_i^\dag f_j+ {h}^2_{ij} f_i^\dag f_j^\dag + \text{H.c.}$ and ${\cal Z}$ is the normalization factor. Similar to the zero-temperature limit, the eigenvalues of $\hat{H}_E$ can be found in terms of generalized Green function which includes the pairing 
\begin{align} \label{eq:Gmat}
G_{ij}=\left(\begin{array}{cc}
1-[C^{\sT}]_{ij} & [F^\dag]_{ij} \\
F_{ij} & C_{ij}
\end{array}
\right),
\end{align}
where 
\begin{align} \label{eq:corr}
C_{ij} &=\braket{f_i^\dag f_j}=  \Tr(e^{-\beta \hat H} f_i^\dag f_j), \\
F_{ij}&=\braket{f_i^\dag f_j^\dag}=  \Tr(e^{-\beta \hat H} f_i^\dag f_j^\dag),
\end{align}
are the thermal two-body correlators and the particle-hole correlators, respectively. 
The rest of the procedure to construct the partially transposed density matrix
is identical to that of the zero-temperature,
as explained in Ref.~\cite{Shap_pTR}

\section{Definition of the theta functions}
\label{app:theta}

The Jacobi theta functions in the main text are given by
\begin{align}
\vartheta_1(z | \tau) &=
2 e^{\pi i \tau/4} \sin (\pi z) \prod_{n=1}^\infty (1-q^n) (1- yq^n) (1-y^{-1} q^n),\\
\vartheta_2(z | \tau) &=
2 e^{\pi i \tau/4} \cos (\pi z) \prod_{n=1}^\infty (1-q^n) (1+ yq^n) (1+y^{-1} q^n),\\
\vartheta_3(z | \tau) &=
\prod_{n=1}^\infty  (1-q^n) (1+y q^{n-1/2})(1+y^{-1} q^{n-1/2}), \\
\vartheta_4(z | \tau) &=
\prod_{n=1}^\infty (1-q^n) (1 - y q^{n-1/2})(1-y^{-1} q^{n-1/2}) \ ,
\end{align}
where $y = e^{2 \pi i z}$ and $q = e^{2 \pi i \tau}$.

\section{Absence of sudden death in a system of two fermionic modes} 
\label{app:suddendeath}

In this appendix, we give a basic example of free fermions on a two-site lattice model and explain why entanglement negativity does not have a sudden death at finite temperatures. As a reference, we compute the bosonic partial transpose of fermionic density matrix and find that the corresponding negativity does show a sudden death. As we will see, the sudden death observed in the latter does not have any physical meaning in the fermionic language and is merely due to the wrong choice of definition.

A two-site lattice model is described by the Hamiltonian
\begin{align}
\hat{H}= - \Delta (f_1^\dag f_2 + f_2^\dag f_1),
\end{align}
where $\Delta$ is the tunneling amplitude.
The Hilbert space is four-dimensional composed of the vacuum state $\ket{00}\equiv\ket{0},$ singly occupied states $\ket{10}=f_1^\dag \ket{0}, \ket{01}=f_2^\dag \ket{0},$ and a doubly occupied state $\ket{11}=f_1^\dag f_2^\dag \ket{0}$. 
The eigen-spectrum is given by
\begin{align}
\ket{\Psi_1} &=\frac{1}{\sqrt{2}}(f_1^\dag+f_2^\dag)\ket{0} ,  &&\epsilon_1=-\Delta, \nonumber \\
\ket{\Psi_2}&=\ket{0}  ,  &&\epsilon_2=0,\nonumber \\
\ket{\Psi_3}&=f_1^\dag f_2^\dag\ket{0} ,  &&\epsilon_3=0, \nonumber\\
\ket{\Psi_4}&=\frac{1}{\sqrt{2}}(f_1^\dag-f_2^\dag)\ket{0} ,  &&\epsilon_4=\Delta.
\end{align}
The thermal density matrix in the basis $\{\ket{00}, \ket{10}, \ket{01}, \ket{11}\}$ can be found easily
\begin{align} \label{eq:thermal_twosite}
\rho= e^{-\beta \hat{H}}=\frac{1}{\cal Z}
\left( \begin{array}{cccc}
1 & 0 & 0 & 0\\
0 &\cosh (\beta\Delta)& \sinh(\beta\Delta)& 0 \\
0& \sinh (\beta\Delta)& \cosh(\beta\Delta)& 0 \\
0& 0 & 0 & 1 
\end{array} \right)
\end{align}
where the normalization factor is ${\cal Z}=2+2\cosh(\beta\Delta)$. We first show the inefficiency of the von-Neumann or R\'enyi entropies at finite temperatures. The reduced density matrix for $A_1$ subsystem is found to be $\rho_{A_1}=\frac{\I}{2}$. Hence, any entanglement measure is independent of temperature, even in high temperature limit. This clearly does not make any sense. Let us now take the bosonic partial transpose (\ref{eq:parttrans_b}) of the density matrix
\begin{align}
\rho^{\sT^b_1}=\frac{1}{\cal Z}
\left( \begin{array}{cccc}
1 & 0 & 0 & \sinh(\beta\Delta) \\
0 &\cosh (\beta\Delta)& 0 & 0 \\
0& 0 & \cosh(\beta\Delta)& 0 \\
\sinh (\beta\Delta) & 0 & 0 & 1 
\end{array} \right).
\end{align}
The singular values $\{\lambda_i\}$ of $\rho^{\sT^b_1}$ are ${\cal Z}^{-1}\{\cosh(\beta\Delta),\cosh(\beta\Delta),1+\sinh(\beta\Delta),1-\sinh(\beta\Delta) \}$. Note that at low temperatures $T\ll \Delta$ and as long as $\lambda_4 <0$, we get non-zero entanglement negativity ${\cal E}>0$. However, above a critical temperature $T>T_{SD}$ all singular values become positive and the entanglement negativity vanishes. This is called a ``sudden death"~\cite{Calabrese_Ft2015,Sherman2016,YeJePark}. In this case, $T_{SD}$ is found to be
\begin{align}
T_{SD}=\frac{\Delta}{\ln (\sqrt{2}+1)}.
\end{align}

Taking the fermionic partial transpose (\ref{eq:app_f_21}) leads to the matrix
\begin{align}
\rho^{\sT_1}=\frac{1}{\cal Z}
\left( \begin{array}{cccc}
1 & 0 & 0 & i\sinh(\beta\Delta) \\
0 &\cosh (\beta\Delta)& 0 & 0 \\
0& 0 & \cosh(\beta\Delta)& 0 \\
i\sinh (\beta\Delta) & 0 & 0 & 1 
\end{array} \right),
\end{align}
with a  four-fold degenerate singular values  
$\lambda=\cosh(\beta\Delta)/{\cal Z}$. Therefore, the entanglement negativity associated with this density matrix becomes ${\cal E}=\ln(4\lambda)$ which asymptotically vanishes at high temperatures with no sudden death.

To understand the sudden suppression of the negativity in the bosonic partial transpose for $T\geq T_{SD}$, let us write an explicit expansion for the density matrix. The general form of density matrix of two qubits is given by~\cite{Ben-Aryeh2015},
\begin{align}
\rho=&\frac{1}{2} \sum_{i=1}^3 |t_i| \left[\frac{(\I-\sigma_i)_{A_1}}{2} \otimes \frac{(\I-\sgn(t_i) \sigma_i)_{A_2}}{2}+
\frac{(\I+\sigma_i)_{A_1}}{2} \otimes \frac{(\I+\sgn(t_i) \sigma_i)_{A_2}}{2} \right] \nonumber \\ 
& + \frac{1}{4} \left(1- \sum_{i=1}^3 |t_i| \right)  \I_{A_1}\otimes \I_{A_2},
\label{eq:2qubit_decomp}
\end{align}
where $t_i\in \mathbb{R}$.
For the purpose of discussion, let us remind ourselves the separability condition of density matrices. A state is called separable, if it can be decomposed as
\begin{align}
\rho_{\text{sep}}= \sum_j w_j \rho_j^1 \otimes \rho_j^2,
\end{align}
where $w_j$ are real positive coefficients and $\rho_j^1$ and $\rho_j^2$ are density matrices acting on the sub-Hilbert spaces ${\cal H}_1$ and ${\cal H}_2$, respectively. It is easy to see that the logarithmic negativity of a separable state is zero.

As mentioned, all the coefficients $w_j$ are positive in a separable state, which in the case of Eq.~(\ref{eq:2qubit_decomp}) implies $\sum_{i=1}^3 |t_i| \leq 1$. Using the density matrix (\ref{eq:thermal_twosite}), we get 
\begin{align}
t_1=t_2=\frac{\sinh(\beta\Delta)}{1+\cosh(\beta\Delta)}, \quad 
t_3=\frac{1-\cosh(\beta\Delta)}{1+\cosh(\beta\Delta)}.
\end{align}
Therefore, the separability condition leads to the inequality 
\begin{align}
2 \sinh(\beta\Delta) + |1-\cosh(\beta\Delta)| \leq 1+ \cosh(\beta\Delta),
\end{align}
which can be simplified into $\sinh(\beta\Delta) \leq 1$. This inequality in turn gives rise to the same inequality $T\geq T_{SD}$ as found earlier. 
This observation means that $\rho$ is separable in the bosonic formalism. However, the density matrices $\I\pm\sigma_x=1\pm (f +f^\dag)$ are not legitimate fermionic density matrices, because they violate fermion-number parity symmetry. Hence, $\rho$ is not separable in the fermionic formalism. 
This example suggests that the absence of sudden death in the fermionic negativity originates from the fermion-number parity constraint on density matrices.


\section*{References}

\bibliography{refs}

\end{document}